\documentclass{aa}      
\newcommand{\todo}[1]{}
\renewcommand{\todo}[1]{{\color{red} TODO: {#1}}}
\usepackage{multirow}
\usepackage{tabularx}
\usepackage{gensymb}
\usepackage{tabularray}
\usepackage{tablefootnote}
\usepackage{footnote}
\usepackage{natbib}
\usepackage{xcolor}
\definecolor{customblue}{HTML}{0073BD}
\usepackage[colorlinks, citecolor=customblue]{hyperref}
\bibpunct{(}{)}{;}{a}{}{,} % to follow the A&A style
\usepackage{graphicx}
\usepackage{txfonts}
\usepackage{stfloats}  % 在导言区添加

\usepackage{tabularx}
\usepackage{soul}
\usepackage[switch]{lineno}

\usepackage{orcidlink}
\usepackage{adjustbox}
\usepackage{subcaption}

\newcommand{\nsample}{55,109}
\newcommand{\nlofar}{27,625}% 样本数
\newcommand{\nboth}{928}
\newcommand{\fagn}{$1.68\pm 0.05\%$}
\newcommand{\flae}{$3.36 \pm 0.11\%$}
\newcommand{\fagnc}{$1.77\pm0.04\%$}
\newcommand{\flaec}{$18.15\pm0.14\%$}

\begin{document} 
   \title{Radio-detected Ly$\alpha$ emitters at $1.88 < z < 3.52$: AGN fraction and Ly$\alpha$ emission}
   \titlerunning{Radio AGN fraction and Ly$\alpha$ visibility}

   \author{Sai Zhai
          \inst{1}\orcidlink{0000-0001-6677-9940}
          \and
          Huub Röttgering\inst{1}\orcidlink{0000-0001-8887-2257}
          \and 
          Anniek J. Gloudemans \inst{2}\orcidlink{0009-0009-8274-441X}
          \and 
           Erin Mentuch Cooper \inst{3,4}\orcidlink{0000-0002-2307-0146}
          \and  
          Maya H. Debski \inst{3}\orcidlink{0000-0002-1998-5677}
           \and
          Gregory Zeimann \inst{5}\orcidlink{0000-0003-2307-0629}
          \and
          Matt J. Jarvis\inst{6,7}\orcidlink{0000-0001-7039-9078}
          \and 
          Leah K. Morabito \inst{8,9}\orcidlink{0000-0003-0487-6651}
          \and
          Donald P. Schneider \inst{10,11} \,\orcidlink{0000-0001-7240-7449}
          \and
          Daniel J. Farrow \inst{12,13}\orcidlink{0000-0003-2575-0652}
          \and 
          Gary J. Hill \inst{4,3}\orcidlink{0000-0001-6717-7685}
          \and 
          Caryl Gronwall \inst{10,11}\orcidlink{0000-0001-6842-2371}
          \and 
          Yuming Fu \inst{1,14}\orcidlink{0000-0002-0759-0504}
          }

   \institute{Leiden Observatory, Leiden University, P.O. Box 9513, 2300 RA Leiden, The Netherlands\\
              \email{szhai@strw.leidenuniv.nl}
        \and 
        NSF NOIRLab, Gemini Observatory, 670 N A'ohoku Place Hilo, HI 96720, USA
         \and
         Department of Astronomy, The University of Texas at Austin, 2515 Speedway Boulevard, Austin, TX 78712, USA
         \and
         McDonald Observatory, The University of Texas at Austin, 2515 Speedway Boulevard, Austin, TX 78712, USA
         \and 
         Hobby Eberly Telescope, University of Texas, Austin, Austin, TX, 78712
         \and 
         Astrophysics, Department of Physics, University of Oxford, Keble Road, Oxford, OX1 3RH, UK
         \and
         Department of Physics and Astronomy, University of the Western Cape, Robert Sobukwe Road, 7535 Bellville, Cape Town, South Africa
         \and
         Centre for Extragalactic Astronomy, Department of Physics, Durham University, South Road, Durham DH1 3LE, UK  
         \and 
         Institute for Computational Cosmology, Department of Physics, Durham University, South Road, Durham DH1 3LE, UK 
         \and
         Department of Astronomy \& Astrophysics, The Pennsylvania
State University, University Park, PA 16802, USA
        \and
Institute for Gravitation and the Cosmos, The Pennsylvania State University, University Park, PA 16802, USA
        \and
         E. A. Milne Centre for Astrophysics
        University of Hull, Cottingham Road, Hull, HU6 7RX, UK
        \and 
        Centre of Excellence for Data Science,
        Artificial Intelligence \& Modelling (DAIM),
        University of Hull, Cottingham Road, Hull, HU6 7RX, UK    
        \and
        Kapteyn Astronomical Institute, University of Groningen, PO Box 800, 9700 AV Groningen, The Netherlands
      }

   \date{Received xx; accepted xx}
 
  \abstract
   {Ly$\alpha$ emitters (LAEs) are galaxies with strong Ly$\alpha$ emission, tracing early star formation and ionizing radiation. Their connection to active galactic nuclei (AGNs) is key to understanding the mechanisms behind (extended) Ly$\alpha$ emission.} 
   {In this work, we measure the fraction of LAEs identified as radio-emitting AGN ($f_{\text{AGN, radio}}$) and the fraction of radio sources that exhibit Ly$\alpha$ emission ($f_{\mathrm{Ly}\alpha}$) to investigate the connection of radio AGN activity to Ly$\alpha$ emission at $1.88 < z < 3.52$.} 
   {We identify  \nboth \ sources that are detected in both the Hobby-Eberly Telescope Dark Energy Experiment (HETDEX) and the LOw Frequency ARray (LOFAR) surveys. These matches are drawn from \nsample \ spectroscopically-confirmed LAEs and \nlofar \  radio sources. }
   {After applying completeness corrections, we obtain $f_{\text{AGN, radio}} =$ \fagnc \ and $f_{\mathrm{Ly}\alpha} =$ \flaec \ (for Ly$\alpha$ fluxes above $3 \times 10^{-17}$ erg s$^{-1}$ cm$^{-2}$ and radio flux densities above $0.1$  mJy at 144 MHz). The fraction $f_{\text{AGN,radio}}$ increases from $0.4\pm0.1 \%$ at $\log_{10} L_{\mathrm{Ly\alpha, erg \ s^{-1}}} \approx 42.6$ to $9.7\pm1.3 \% $ at $\log_{10} L_{\mathrm{Ly\alpha, erg \ s^{-1}}} \approx 44.5$. The fraction $f_{\mathrm{Ly\alpha}}$ rises from $0.7\pm0.1\%$  at $\log_{10} L_{150,\rm W \ Hz^{-1}} \approx 24.6$ to $55.8\pm14.5 \%$ for $\log_{10} L_{150,\rm W \ Hz^{-1}} \approx 28.3$.
   `LAEs with radio AGN' and `optical AGN with Ly$\alpha$ emission' exhibit similar radio luminosity distributions 
above the AGN threshold ($L_{150\,\rm MHz} > 10^{23.9}$ W Hz$^{-1}$), 
though optical AGN show systematically higher Ly$\alpha$ luminosities. 
We find no statistically significant correlations of Ly$\alpha$ luminosity with both radio 
luminosity and low-frequency spectral index, suggesting that Ly$\alpha$ emission is regulated by gas properties rather than jet power alone.
  We also find a positive correlation between the full width at half maximum (FWHM) and luminosity of the Ly$\alpha$ line, suggesting that more luminous sources have broader lines, potentially due to enhanced outflows or velocity dispersions. However, we find no correlation between FWHM and radio size, indicating minimal direct jet-gas interaction in these systems.}
   {Our results show that most Ly$\alpha$ emission at $1.88<z<3.52$ is powered by star formation, with radio AGN activity confined to a small luminous subset (\fagnc). While luminous systems preferentially host radio AGN, the lack of correlation between Ly$\alpha$ and radio luminosities in individual sources indicates that Ly$\alpha$ emission is governed primarily by host galaxy gas properties rather than direct AGN jet coupling.}


   \maketitle
%
%-------------------------------------------------------------------

\section{Introduction}
 
Lyman-alpha emitters (LAEs) are a population of galaxies identified by their strong Ly$\alpha$ emission line at a rest-frame wavelength of 1216~\AA\ \citep{Partridge1967ApJ...147..868P}. Their prominent $\rm Ly \alpha$ emission line profiles make them valuable tracers of early galaxy formation and evolution (e.g., \citealt{Hu1996Natur.382..231H, Pascarelle1996Natur.383...45P, 2020ARA&A..58..617O, 2021MNRAS.505.1382M}), as well as useful probes of large scale environments such as galaxy clusters \citep{Orsi2008MNRAS.391.1589O,Venemans2005A&A...431..793V}.

Although LAEs were initially thought to be primarily low-mass, metal-poor star-forming galaxies \citep{Gawiser2007ApJ...671..278G,Guaita2011ApJ...733..114G}, subsequent studies have revealed a more diverse population, including massive and dusty galaxies \citep{Finkelstein2009ApJ...691..465F,Shimizu2011MNRAS.418.2273S,Bridge2013ApJ...769...91B}, and systems hosting AGNs \citep{Pentericci2009A&A...494..553P, Kornei2010ApJ...711..693K, Gloudemans2021A&A...648A...7G}. The origins of Ly$\alpha$ emission is therefore various: star formation in young O/B stars can ionize interstellar medium \citep{Hui1997MNRAS.292...27H,Dijkstra_2014}, while AGNs can also ionize surrounding gas \citep{Smith2007MNRAS.378L..49S,Cantalupo2014Natur.506...63C,Borisova2016ApJ...831...39B,Costa2022MNRAS.517.1767C}. Outflows, gravitational cooling radiation, major merger and fluorescence could also play a significant role (\citealt{Rosdahl2012MNRAS.423..344R,Kusakabe2022A&A...660A..44K}). Studying each of these contributions can shed light on how galaxies acquire, lose, or recycle gas, and how supermassive black holes might affect or be influenced by these processes (e.g., \citealt{2025ApJ...982...54B}).

Over the past two decades, tens of thousands of LAEs have been identified through two primary detection strategies: narrow-band (NB) imaging and untargeted spectroscopic surveys (e.g., \citealt{Ouchi2008ApJS..176..301O, Bacon2010SPIE.7735E..08B, Adams2011ApJS..192....5A, Konno2018PASJ...70S..16K, Gebhardt2021ApJ...923..217G}). The NB imaging technique detects excess flux in a narrow filter relative to broadband photometry at a specific redshift and has been widely used with large-field cameras such as Suprime-Cam and Hyper Suprime-Cam (HSC, \citealt{Miyazaki2012SPIE.8446E..0ZM, Ciardullo2012ApJ...744..110C,2020ARA&A..58..617O}). Meanwhile, untargeted spectroscopic surveys, such as those conducted by the Multi-Unit Spectroscopic Explorer (MUSE, \citealt{Bacon2010SPIE.7735E..08B}) and the Hobby-Eberly Telescope Dark Energy Experiment (HETDEX, \citealt{Gebhardt2021ApJ...923..217G,Hill2021AJ....162..298H}), allow for the detection of more reliable LAE populations by directly measuring emission-line spectra over wide areas.

Beyond optical surveys, radio observations provide a dust-unbiased and complementary view of LAE physics (e.g., \citealt{Padovani2016A&ARv..24...13P}). Radio continuum can trace obscured star formation and reveal AGN activity via synchrotron emission from jets and lobes (e.g., \citealt{Heckman2014ARA&A..52..589H}). Radio power and morphology offer diagnostics independent of resonant Ly$\alpha$ radiative transfer and provide insight on feedback and environmental interactions (e.g., \citealt{Hardcastle2020NewAR..8801539H}). Combining radio and Ly$\alpha$ data can disentangle the mechanisms powering Ly$\alpha$ emission.

An important topic in this context is determining the fraction of LAEs that host AGNs and, conversely, the fraction of radio sources with Ly$\alpha$ emission. Quantifying these fractions is crucial for understanding the diverse energy source of Ly$\alpha$ emission, the role of AGN in galaxy evolution, and how Ly$\alpha$ emitting and radio populations trace different environments in high-redshift environments (e.g., \citealt{Ouchi2008ApJS..176..301O,Konno2018PASJ...70S..16K}). Previous research on optically and/or X-ray detected AGN suggests that the AGN fraction among LAEs grows with Ly$\alpha$ luminosity (e.g., \citealt{Wold2014ApJ...783..119W,Matthee2017MNRAS.471..629M,Sobral2018MNRAS.477.2817S,Calhau2020MNRAS.493.3341C}). Compared to X-ray or optically selected AGNs, radio-selected AGNs provide a dust-unbiased and complementary view of AGN activity. At intermediate to high redshift, radio-loud AGNs can serve as effective tracers of supermassive black hole growth \citep{Merloni2003MNRAS.345.1057M,Padovani2016A&ARv..24...13P,Delvecchio2017A&A...602A...3D,Hickox2018ARA&A..56..625H,Wang2024A&A...689A.327W}, although the limited depth of large-area radio surveys means that only the most powerful AGNs are typically observed \citep{Condon1998AJ....115.1693C,Best2005MNRAS.362....9B,Shimwell2019A&A...622A...1S}. Furthermore, several studies have investigated potential links between Ly$\alpha$ emission and radio power in radio-loud AGNs at $z > 4$ \citep{Saxena2017MNRAS.469.4083S, Gloudemans2021A&A...648A...7G}. These efforts, however, are based on limited samples, and a comprehensive understanding of the connection between radio and Ly$\alpha$ emission remains elusive. In addition, most previous work on Ly$\alpha$ fractions has focused on Lyman break galaxies \citep{Stark2010MNRAS.408.1628S,Stark2011ApJ...728L...2S,Schenker2012ApJ...744..179S}. Extending such analyses to a larger sample of radio-selected sources offers a promising approach to better understand the connection between AGN and Ly$\alpha$ emission, in line with the goals of the William Herschel Telescope Enhanced Area Velocity Explorer (WEAVE)–LOw Frequency ARray (LOFAR) survey \citep{vanHaarlem2013A&A...556A...2V,Smith2016sf2a.conf..271S}.

The LOFAR Two-metre Sky Survey (LoTSS) provides wide-area, deep, and relatively high-resolution ($6''$) imaging at 120–168 MHz, with excellent sensitivity (down to $70 \ \mu \rm Jy \ beam^{-1}$) to steep-spectrum AGN \citep{Shimwell2019A&A...622A...1S}. Its sky coverage overlaps with the spectroscopic LAE surveys in the HETDEX Spring field (see Fig. 2 in \citealt{Shimwell2019A&A...622A...1S}), which substantially increases the number of LAEs at $1.88<z<3.52$ \citep{MentuchCooper2023ApJ...943..177M}. \citet{Debski2025ApJ...978..101D} construct the HETDEX–LOFAR spectroscopic-redshift catalogue, yielding robust redshift for 9710 radio sources (including 9087 new measurements), enabling a systematic assessment of LAE–radio connections.

This work investigates the connection between Ly$\alpha$ emission and radio AGN activity at $z\sim2$--3 using the HETDEX--LOFAR spectroscopic redshift catalogue. We address two complementary questions: (1) what fraction of LAEs host radio-detected AGNs, and (2) what fraction of radio sources exhibit Ly$\alpha$ emission. These measurements allow us to assess both the contribution of radio AGNs to LAE samples and the prevalence of Ly$\alpha$ emission among radio-selected AGNs. The main goal is therefore to quantify the overlap between LAEs and radio AGNs and to clarify how this overlap affects the interpretation of both populations. Section \ref{data} presents the data we used and the sample selection. Section \ref{result} shows the two important fraction mentioned above and the optical and radio properties of LAEs and AGNs. Section \ref{discussion} discusses the reliability and implications of our findings. Section \ref{conclusion} summarizes our results. We assume a flat cosmological model with $ H_{0} = 70 \  \rm km \ s^{-1} \ Mpc^{-1}, \Omega_{m} = 0.3$, and $\Omega_{\Lambda} = 0.7$ \citep{PlanckCollaboration2016A&A...594A..13P}, and use celestial coordinates in the International Celestial Reference System (ICRS) frame.

\section{Data}\label{data}

\subsection{HETDEX optical data}

HETDEX is a untargeted integral field spectroscopy survey aimed at detecting more than a million LAEs over 540 $\rm deg^{2}$ within $1.88<z<3.52$ using the Hobby-Eberly Telescope (HET, \citealt{Ramsey1998SPIE.3352...34R,Hill2021AJ....162..298H}). HETDEX utilizes the Visible Integral-field Replicable Unit Spectrograph (VIRUS, \citealt{Hill2021AJ....162..298H}), which gathers spectra from $\sim$35,000 fibres simultaneously into 156 spectrographs. The spectral wavelength coverage of VIRUS spans 3500–5500 $\AA$,  with a resolution of $R \sim 800$ \citep{Hill2021AJ....162..298H, Davis2023ApJ...954..209D}. In HETDEX, LAEs are defined as galaxies with Ly$\alpha$ emission for which HETDEX does not measure AGN activity \citep{Davis2023ApJ...954..209D}.  The majority of these LAEs ($\geq 96\%$) are classical LAEs which have rest frame equivalent width $\rm EW_{0} \geq 20 \AA$ \citep{2020ARA&A..58..617O}. AGN in HETDEX are identified either through the presence of at least two significant AGN lines (e.g., Ly$\alpha$ and C$\rm IV$ $\lambda1549$ line pair) or through the detection of a single broad emission line with FWHM $>$ 1000 $\rm km \ s^{-1}$ (e.g., Ly$\alpha$, \citealt{Liu_sample2022ApJS..261...24L}). Previous work concluded that most LAEs observed in the HETDEX redshift window are compact, low-metallicity, and rapidly star-forming galaxies \citep{Hagen2016ApJ...817...79H,Davis2023ApJ...954..209D}. 

This work uses the internal fourth data release of HETDEX (HDR4), which includes 6 years of observations from August 2017 to August 2023, covering 67.48 $\rm deg^{2}$ of the sky \citep{MentuchCooper2023ApJ...943..177M,Debski2025ApJ...978..101D}. This catalogue includes 920,715 LAE candidates in the redshift range $1.88 <z< 3.52$, with SNR $> 4.8$. This threshold was adopted by the HETDEX catalogue because, at this limit, the survey meets the required source density specification for emission-line detections, including LAEs \citep{Gebhardt2021ApJ...923..217G,MentuchCooper2023ApJ...943..177M,House2023ApJ...950...82H}. This catalogue also provides key measurements including
the line widths, fluxes, wavelengths, and emission line luminosities.

\subsection{HETDEX-LOFAR spectroscopic redshift catalogue}\label{catalog1}

We additionally use the HETDEX-LOFAR spectroscopic redshift catalogue of \cite{Debski2025ApJ...978..101D}, which was constructed  by cross-matching sources from the LOFAR Two-metre Sky Survey (LoTSS, \citealt{Shimwell2019A&A...622A...1S}) DR1 with the HETDEX DR4 dataset within SPRING footprint. In total, \cite{Debski2025ApJ...978..101D} obtained targeted VIRUS spectroscopy for 28,705 LoTSS DR1 sources within the HETDEX DR4 Spring field. Of these, 7409 sources have spectroscopic counterparts within 
5\arcsec\ in the HDR4  catalogue. The vast majority of sources in this catalogue have continuum S/N $< 2$ per 2\,\AA\ pixel in the wavelength range 4670--4870\,\AA. Spectroscopic redshift were derived using Diagnose \citep{Debski2024zndo..13755510D}, a spectral classification code, in combination with the HETDEX HDR4 internal catalogue spectroscopic redshift as described in \citet{MentuchCooper2023ApJ...943..177M} and additional archival data. In total, redshift were obtained for 9710 sources, of which 9087 had no previously reported spectroscopic redshift.

HETDEX source classification follows a sequential procedure where all `AGNs' are excluded prior to `LAE' classification. As a result, the `LAE' sample consists entirely of non-AGN LAEs, which we refer to as `star-forming LAEs'. In addition, some sources identified as `AGNs' do exhibit Ly$\alpha$ emission; we refer to these as `optical AGN with Ly$\alpha$ emission'. Together, we group these two populations as 'Ly$\alpha$ emitting sources'. In total, this radio-selected sources catalogue  contains 1075 `star-forming LAEs' and 804 `optical AGN with Ly$\alpha$ emission'.

We further refine the source classification using the HETDEX DR4 AGN catalogue of \cite{liu2025ApJS..276...72L}, which employs multi-Gaussian fitting for emission lines to provide improved spectral decomposition. This catalogue incorporates optical AGN classifications that were not yet available in \cite{Debski2025ApJ...978..101D}. We update our source classifications accordingly using this catalogue.

\begin{figure}
    \centering
    \includegraphics[width=\linewidth]{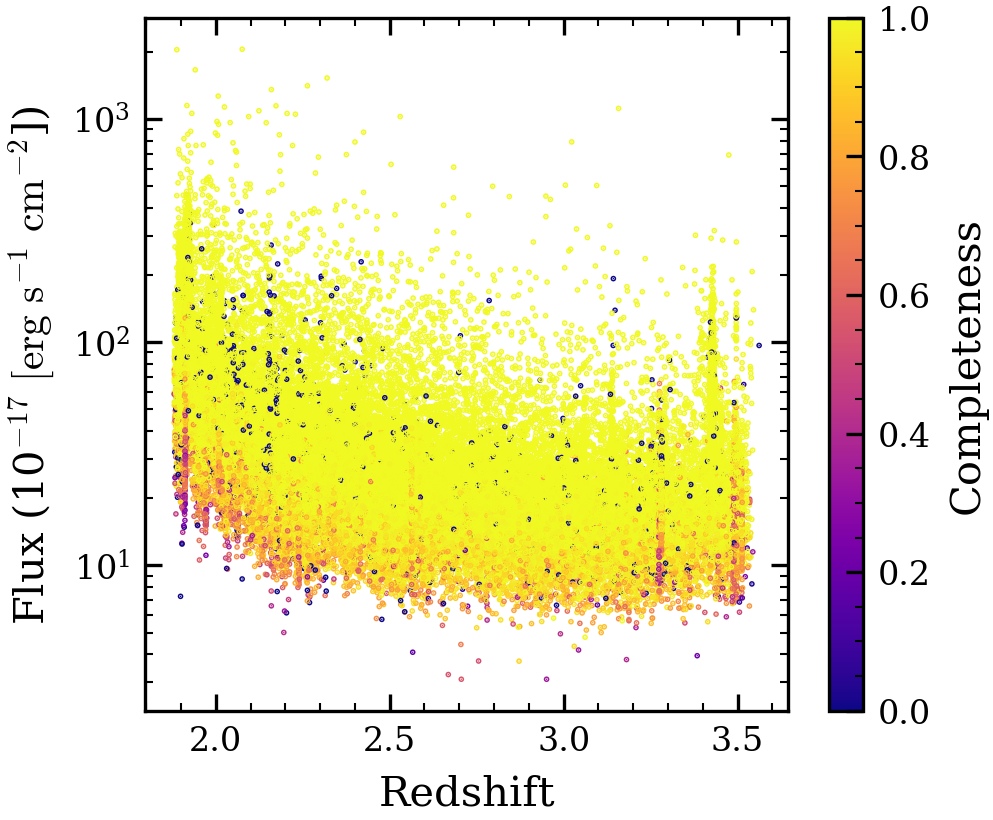}
    \caption{Ly$\alpha$ line flux  as a function of redshift. The colour-bar represents the completeness obtained through the HETDEX API \protect\footnotemark{} for \nsample \ LAEs, selected as discussed in the text. }
    \label{fig:completeness}
\end{figure}
\footnotetext{HETDEX API access portal: \url{https://hetdex-api.readthedocs.io/en/latest/}}

\subsubsection{\texorpdfstring{Ly$\alpha$ luminosity}{Ly alpha luminosity}}

We convert the Ly$\alpha$ flux  to Ly$\alpha$ luminosity by using:
\begin{equation}
     L_{\rm Ly\alpha} = 4 \pi d_{\rm L}^{2}f_{\rm Ly\alpha},
\end{equation}
where $L_{\rm  Ly\alpha}$, $d_{\rm  L}$, and $f_{\rm Ly\alpha}$ represent the Ly$\alpha$ luminosity, the luminosity distance, and the Ly$\alpha$ flux, respectively. We adopt the $f_{\rm Ly\alpha}$ measurements reported by \cite{Erin_inprep} for extended sources. In their study, each object is classified as either compact or extended based on a fit to the Ly$\alpha$ spatial flux distribution.
Sources are classified as extended if a two-component model (point-like core plus exponential halo) provides a better fit, and as compact if a single point-source model fits best (see \citealt{Erin_inprep} for details). The total Ly$\alpha$ flux are obtained by integrating the surface-brightness radial profiles for extended sources. For compact sources, we use Ly$\alpha$ flux from the HETDEX DR4 catalogue.

\subsubsection{LOFAR radio data}
As mentioned in Sect. \ref{catalog1}, \cite{Debski2025ApJ...978..101D} built the HETDEX-LOFAR spectroscopic redshift catalogue using radio data from the LoTSS DR1 \citep{Shimwell2019A&A...622A...1S}, which already fully covers the HETDEX SPRING footprint. This HETDEX-LOFAR catalogue defines the parent radio-source sample used in this work.

LoTSS DR2 \citep{Shimwell2022yCat..36590001S} expands the sky coverage beyond that of DR1 and provides additional source information. In this work, we use LoTSS DR2 only for supplementary measurements. First, when estimating the expected fraction of radio sources in the redshift range (1.88 < z < 3.52) in Section \ref{fradio}, we use photometric redshift from the LOFAR deep field, which only covered by DR2. Second, we adopt the radio size measurements from \cite{Hardcastle2023A&A...678A.151H} in Section \ref{size}, which are not available in LoTSS DR1, to characterize the physical extent of our radio sources.  These DR2-based quantities are not used to redefine the parent radio-source sample.

\subsection{Sample selection: LAEs with optical and radio AGN} \label{sample}

We construct our sample from the HETDEX–LOFAR spectroscopic catalogue presented by \citet{Debski2025ApJ...978..101D} (see Sect.~\ref{catalog1} for a brief overview). We first select reliable Ly$\alpha$ emitting sources in the HETDEX DR4 Spring field with Ly$\alpha$ line SNR $> 6.5$ and detection confidence level $> 0.5$, following the method described in \cite{Lujan2022ApJ...929...90L}. This approach yields a sample of \nsample \ secure Ly$\alpha$ emitting sources from the full HETDEX DR4 catalogue, which are used as the parent population for the subsequent comparisons. 

We then consider the radio-detected sources in the same field. To ensure consistency with the HETDEX parent sample, we apply the same cuts (Ly$\alpha$ line $\rm SNR > 6.5$ and detection confidence $> 0.5$) to the cross-matched sample. This process results in a final set of \nboth \ robustly detected sources—comprising both `star-forming LAEs' (514 sources) and `optical AGN with Ly$\alpha$ emission' (414 sources) —all with reliable Ly$\alpha$ emission in HETDEX DR4 and corresponding radio counterparts in LoTSS DR1. A summary of the sample's properties is provided in Table~\ref{tab:1}.

\subsubsection{AGN selection}\label{AGNselection}

Part of this work aims to study the fraction of Ly$\alpha$ emitting sources that host radio-detected AGN. As summarized by \citet{Hardcastle2019A&A...622A..12H}, AGNs selected at different wavelengths often represent overlapping but different populations. Radiatively efficient AGN, typically identified through X-ray, optical, or mid-infrared (MIR) diagnostics, are often hosted by star-forming galaxies where radio emission is weak and star formation dominated \citep[e.g.,][]{Kauffmann2003MNRAS.346.1055K, Brandt2005ARA&A..43..827B, CalistroRivera2016ApJ...833...98C, Mingo2016MNRAS.462.2631M,Gurkan2018MNRAS.475.3010G}. In contrast, radio-loud AGN (RLAGN) exhibit strong radio emission that dominates over their emission at other wavelengths due to strong jet-related emission, even though they can also be luminous in the optical or infrared bands \citep[e.g.,][]{Best2012MNRAS.421.1569B, Heckman2014ARA&A..52..589H}.
MIR colour–colour selection method identifies luminous AGN but may miss lower-luminosity AGN or high-redshift sources, which are often the hosts of radio-emitting AGN \citep{Jarrett2011ApJ...735..112J,Gurkan2014MNRAS.438.1149G, Rovilos2014MNRAS.438..494R, Secrest2015ApJS..221...12S}.

Therefore, in addition to AGNs classified by emission-line diagnostics, we use radio data to assess whether any of the `star-forming LAEs' may in fact host AGN activity. We examine the 150 MHz radio luminosities of all `star-forming LAEs' detected by LOFAR. Most of these sources lie at $\log_{10}(L_{150}/{\rm W\,Hz^{-1}})>24$, i.e. above the luminosity range over which the empirical $L_{150}$--SFR relation for star-forming galaxies was calibrated \citep{Gurkan2018MNRAS.475.3010G}.

The high radio luminosities themselves indicate that these sources are unlikely to be normal star-forming galaxies. If their radio emission were powered purely by star formation, they would require extremely high SFR, well beyond those expected for normal star-forming galaxies and comparable only to rare starburst systems \citep{Kennicutt1998ARA&A..36..189K, Elbaz2011A&A...533A.119E, Rodighiero2011ApJ...739L..40R}. Therefore, rather than treating the inferred SFR as quantitative measurements, we use this comparison to show that star formation alone is unlikely to dominate the observed radio emission.

We therefore conclude that the radio emission in these sources is most plausibly dominated by AGN activity, since AGNs can generate strong low-frequency radio emission through synchrotron radiation from relativistic jets \citep{Hardcastle2019A&A...622A..12H, Smol2017A&A...602A...1S}. We classify these sources as `LAEs with radio AGNs'.

\subsubsection{Selection function}\label{selection_function}

The completeness of Ly$\alpha$ emitting sources in HETDEX is estimated using simulations designed to evaluate the performance of the HETDEX detection algorithm. Specifically, artificial emission lines are generated with input parameters such as observed wavelength ($\rm \lambda_{obs}$), fluxes($f_{\rm Ly\alpha}$), line width ($\rm \sigma_{obs}$), sky position (RA, DEC), and the SNR, and are then injected into the data to test whether the algorithm can recover them \citep{Gebhardt2021ApJ...923..217G,Liu2022ApJ...935..132L}.

The detection completeness is  expressed as a function of these parameters: \begin{equation}
    \rm C_{HETDEX,Lya} = C(\lambda_{obs}, \textit{f}_{\rm Ly\alpha}, \sigma_{obs},RA, Dec, SNR),
\end{equation}
where each variable corresponds to the inputs defined above (Farrow et al. in prep.). The completeness values of all (\nsample) LAEs used in this study are shown in Fig. \ref{fig:completeness}. The lower completeness at $z \sim 2.0 -2.3$  arises from decreased instrumental sensitivity near strong sky lines \citep{MentuchCooper2023ApJ...943..177M}.

%--------------------------------------------------------------------
\section{Results}\label{result}

\subsection{Radio AGN fraction of the Ly$\alpha$ emitting sources}\label{fagn_lya}

\begin{figure}
    \centering
    \includegraphics[width=\linewidth]{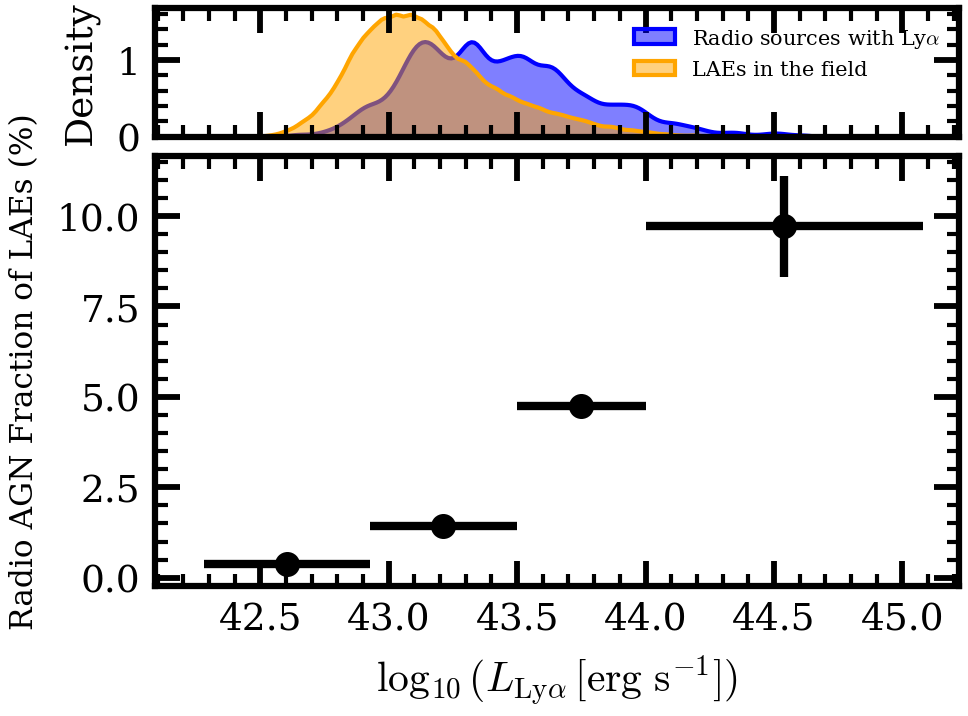}
    \caption{The top panel shows the distribution of “radio sources with Ly$\alpha$ emission” (blue) and “field LAEs” (yellow). The bottom panel presents the fraction of LAEs with radio emission as a function of Ly$\alpha$ luminosity. 
    The fraction of radio-detected LAEs rises with Ly$\alpha$ luminosity.}%Based on the \textbf{dominator} lya range.}
    \label{fig:f_Llya}
\end{figure}

We first investigate the radio AGN fraction of Ly$\alpha$ emitting sources, which we denote as $f_{\rm radio AGN}$. As discussed in Section \ref{AGNselection}, the subset of `star-forming LAEs' detected in both HETDEX and LOFAR are classified as `LAE with radio AGN' due to their potential AGN nature. 
 This fraction is calculated as the ratio of the number of radio AGN ($N_{\rm matched} = N_{\rm LAE\text{-}radio \ AGN} + N_{\rm optical \ AGN\ with \ Ly\alpha \ emission}$) to the total number of Ly$\alpha$ emitting sources in HETDEX ($N_{\rm LAE}$). As mentioned in Section \ref{sample}, HDR4 contains \nsample \ high-quality Ly$\alpha$ emitting sources (SNR > 6.5 and Ly$\alpha$ confidence > 0.5) in the redshift range $1.88<z<3.52$. Among these sources, there are \nboth \ radio AGNs from LoTSS DR1, resulting in an observed radio-AGN fraction of  \fagn \ (Binomial error). 

We correct this fraction for the effect of incompleteness. Section \ref{selection_function} explains how the completeness of HETDEX is estimated in fixed SNR, wavelength, line width, flux  of emission lines, and sky location. The observed $N_{\rm matched}$ is also affected by detection completeness of LOFAR survey. For LOFAR, we apply an average completeness of 0.9 ($C_{\rm LOFAR,ave.}$, with negligible Poisson error, \citealt{Shimwell2019A&A...622A...1S}). The corrected $f_{\rm radio AGN,corr.}$ is determined through:
\begin{equation}
    f_{\rm radio AGN,corr.} =  \frac{ \left( \displaystyle \sum_{i=1}^{N_{\rm matched}} \frac{1}{C_{\rm HETDEX,\textit{i}}} \right)\times \frac{1}{C_{\rm LOFAR,ave.}}}{\displaystyle \sum_{i=1}^{N_{\rm LAE}}{\frac{1}{C_{\rm HETDEX,\textit{i}}}}} ,
\end{equation}
where $C_{\rm HETDEX}$ and $C_{\rm LOFAR,ave.}$ are the completeness of HETDEX and LOFAR, respectively. After applying this correction, the completeness-adjusted radio AGN fraction of Ly$\alpha$ emitting sources increases to \fagnc \ (Uncertainties are estimated using bootstrap resampling with 10,000 iterations. For each bootstrap realization, we resample 100\% of the original sample size with replacement for both the numerator and denominator samples.).

This value is broadly consistent with previous results. For example, \cite{Gawiser2007ApJ...671..278G} reported a X-ray AGN fraction of $1\%$ at $z\sim3.1$, and \cite{Calhau2020MNRAS.493.3341C} found  a radio AGN fraction of $3.2\pm0.3\%$ at $2.2<z<6$. These findings confirm the low incidence of radio emission among Ly$\alpha$-emitting sources and emphasize the rarity of radio AGNs in such populations at $z \approx 2 - 3$.

In addition, Fig.~\ref{fig:f_Llya} shows the radio AGN fraction of Ly$\alpha$-emitting sources ($f_{\rm radio\,AGN}$) as a function of Ly$\alpha$ luminosity. The top panel displays the luminosity distributions of radio sources with Ly$\alpha$ emission and field LAEs, showing that radio-detected sources preferentially occupy the high-luminosity tail. The bottom panel presents a clear positive correlation between the radio AGN fraction and Ly$\alpha$ luminosity: $f_{\rm radio\,AGN}$ increases from $\sim 0.4 \pm 0.1\%$ at $\log_{10}(L_{\rm Ly\alpha}) \approx 42.6$  to $\sim 9.7 \pm 1.3\%$ at $\log_{10}(L_{\rm Ly\alpha}) \approx 44.5$. This trend indicates that more luminous Ly$\alpha$ emitters are significantly more likely to host radio AGNs, consistent with AGN activity contributing to the enhanced Ly$\alpha$ emission.

This luminosity-dependent radio fraction is consistent with previous findings that AGN activity can enhance Ly$\alpha$ emission through mechanisms such as ionizing radiation from the AGN, shocks from radio jets, or increased gas densities in the circumnuclear region (e.g., \citealt{Heckman1991ApJ...370...78H, Cai2018ApJ...861L...3C, Calhau2020MNRAS.493.3341C, Gonz2023A&A...679A..41G}).

\begin{table*}[htbp]
\caption{Observed and completeness-corrected fractions of Ly$\alpha$-emitting and radio-selected sources.}
\label{tab:1}
\centering
\begin{tabular}{lcc}
\hline
\hline
Type & Observed Value & Completeness-corrected Value \\
\hline
Ly$\alpha$-emitting sources (SNR > 6.5 \& Ly$\alpha$ confidence > 0.5) & \nsample & - \\
Radio sources & \nlofar & - \\
Radio sources with Ly$\alpha$ emission & \nboth & - \\
Radio AGN fraction of Ly$\alpha$-emitting sources & \fagn & \fagnc \\
Ly$\alpha$ emission fraction of radio sources & \flae & \flaec \\
\hline
\end{tabular}
\tablefoot{
The table summarizes the numbers of Ly$\alpha$ emitters, radio sources, and matched sources in the overlap region, together with the observed and completeness-corrected fractions. Completeness corrections account for the incompleteness of both the HETDEX Ly$\alpha$ emitter sample and the LoTSS radio catalogue.
}
\end{table*}

\subsection{Fraction of radio galaxies with Ly$\alpha$ emission}\label{fradio}

We also investigate the fraction of radio sources that exhibit Ly$\alpha$ emission ($f_{\rm Ly\alpha}$). This fraction is defined as the number of matched radio sources with Ly$\alpha$ ($N_{\rm matched}$) emission divided by the number of radio sources in the HETDEX field ($N_{\rm LOFAR}$).

 Ly$\alpha$ emitting sources from HETDEX are specifically selected within the redshift range $1.88 < z < 3.52$. However, radio sources in LoTSS DR1 span a much broader redshift distribution, from $z = 0$ up to $z = 7$, based on both spectroscopic and photometric redshift estimates (e.g., \citealt{Duncan2019A&A...622A...3D, Bhardwaj2024A&A...692A...2B}). To quantify the number of radio sources that lie within the same redshift range as the Ly$\alpha$ emitting sources, we estimate the fraction of LoTSS radio sources that fall in the interval $1.88 < z < 3.52$.

For this purpose, we use the LoTSS DR2 catalogue and the deep field catalogue from the European Large Area ISO Survey-North 1 (ELAIS-N1) field \citep{Shimwell2022yCat..36590001S, 2021A&A...648A...2S}. We cross-match the two catalogues using a matching radius of $3''$ to identify radio sources located in the ELAIS-N1 field and within the sensitivity limits of LoTSS DR2. Based on this cross-match, we find that 18\% of the LoTSS DR2 radio sources fall within the photometric redshift range $1.88 < z < 3.52$. 

Then the corrected $f_{\rm Ly\alpha}$ is measured by:
\begin{equation}
    f_{\rm Ly\alpha,corr.} = \frac{\displaystyle \sum_{i=1}^{N_{\rm matched}} \frac{1}{C_{\rm HETDEX,\textit{i}}} }{N_{\rm LOFAR} \times f_{\rm z,2-3}} ,
\end{equation}
where $N_{\rm matched}$, $C_{\rm HETDEX}$, $N_{\rm LOFAR}$ and $f_{\rm z,2-3}$ are the number of matched sources, completeness of HETDEX, the number of sources in LOFAR-HETDEX Spring field, the estimated fraction of radio sources with $1.88 < z< 3.52$ from LOFAR ELAIS-N1 field. After applying this redshift correction, and also the completeness of HETDEX, the adjusted fraction of radio galaxies exhibiting Ly$\alpha$ emission is \flaec \ (Bootstrap error), suggesting that Ly$\alpha$ emission among high-redshift radio galaxies is fairly common at $z=1.88$–$3.52$. The high Ly$\alpha$ fraction observed among radio-selected sources suggests that such populations are frequently associated with strong ionizing activity, such as AGN. This result is consistent with the findings of \citet{Best2023MNRAS.523.1729B} and \citet{Das2024MNRAS.531..977D}, who reported that the AGN fraction increases with radio luminosity above the minimal values probed in this work ($4.73 \times 10^{24} \  \rm W \ Hz^{-1}$), while the fraction of star-forming galaxies correspondingly decreases.

\begin{figure}
    \centering
    \includegraphics[width=\linewidth]{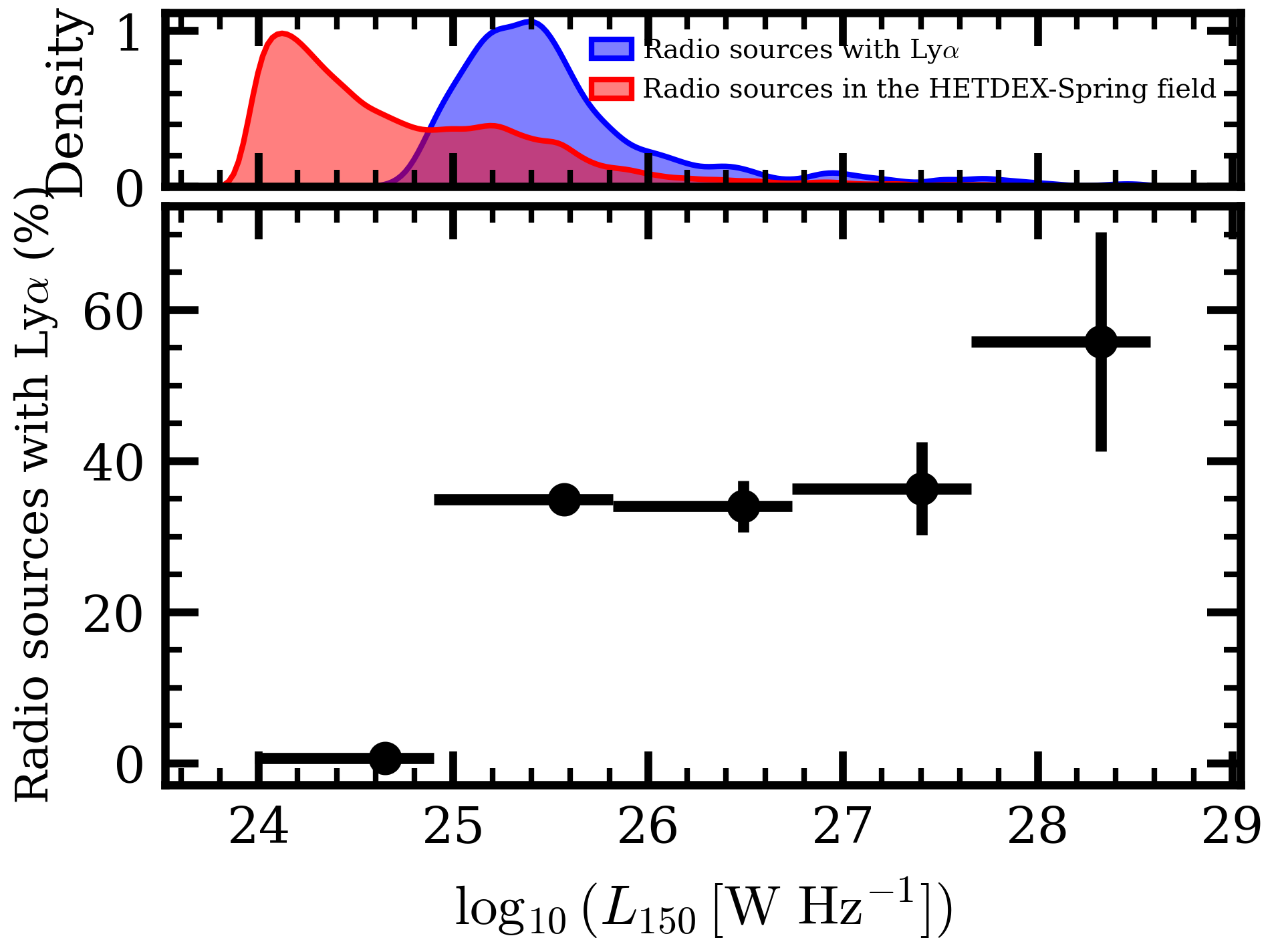}
        \caption{The top panel shows the distribution of “radio sources with Ly$\alpha$ emission” (blue) and “field radio sources” (red). The bottom panel presents the fraction of radio sources with Ly$\alpha$ emission as a function of 150 MHz radio luminosity. A constant trend is observed for radio luminosities greater than $10^{25} \ \rm W \ Hz^{-1}$, suggesting that the fraction remains stable at the highest radio luminosity.} %Based on the min of numberator and max of denomirator.
    \label{fig:f_L150}
\end{figure}

Figure~\ref{fig:f_L150} presents the fraction of radio sources exhibiting Ly$\alpha$ emission ($f_{\rm Ly\alpha}$) as a function of 150 MHz radio luminosity. The top panel compares the luminosity distributions of radio sources with Ly$\alpha$ emission and all field radio sources, showing that Ly$\alpha$-emitting radio sources are predominantly found at higher radio luminosities. 

 At the lowest radio luminosity bin ($\log_{10}(L_{150}) \sim 24.6$), the Ly$\alpha$ detection fraction is very low ($\sim 0.7 \pm 0.1\%$). This is likely driven by selection effects: although star-forming galaxies can exhibit Ly$\alpha$ emission, the faint radio population is dominated by such systems \citep{Hardcastle2019A&A...622A..12H, Best2023MNRAS.523.1729B}, whose Ly$\alpha$ luminosities are typically lower. As a result, wide-field surveys such as HETDEX are preferentially sensitive to the most luminous Ly$\alpha$ emitters, which are more commonly associated with AGN. At higher radio luminosities ($\log_{10}(L_{150}) \gtrsim 25$), where the radio population transitions to AGN-dominated sources \citep{Best2023MNRAS.523.1729B}, the Ly$\alpha$ detection fraction rises substantially to $\sim 35\%$ and remains relatively constant across the range $\log_{10}(L_{150}) \sim 25$--$27$. At the highest luminosities ($\log_{10}(L_{150}) \sim 28.3$), the fraction increases further to $\sim 55.8 \pm 14.5\%$, though the large uncertainty reflects the small number of sources at these extreme luminosities.

We note that our flux-limited sample introduces a redshift-luminosity degeneracy: at higher redshift, only the most luminous sources are detectable. This selection effect is discussed further in Section~\ref{f2}.

\subsection{Optical and radio properties}

In this Section, we study the relation between optical and radio properties of `LAE with radio AGN' and `optical AGN with Ly$\alpha$ emission' (following the classification of HETDEX), such as Ly$\alpha$ luminosities, FWHM, 150 MHz radio luminosities, and radio sizes. 
\begin{figure*}
    \centering
    % --- top row ---
    \begin{subfigure}{0.45\textwidth}
        \centering
        \includegraphics[width=\linewidth]{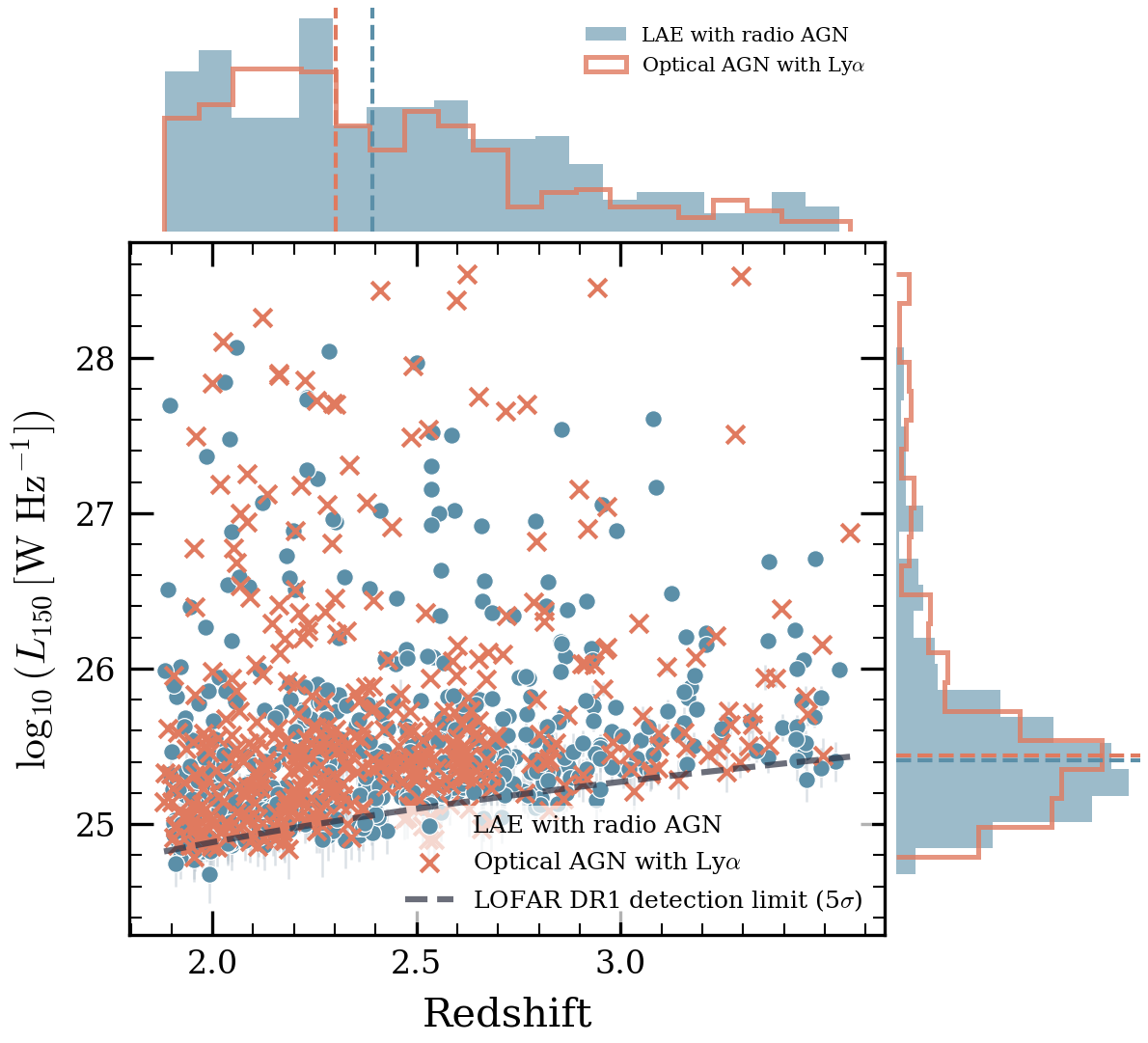}
        \caption{}
        \label{fig:Lradio_z}
    \end{subfigure}
    \hfill
    \begin{subfigure}{0.45\textwidth}
        \centering
        \includegraphics[width=\linewidth]{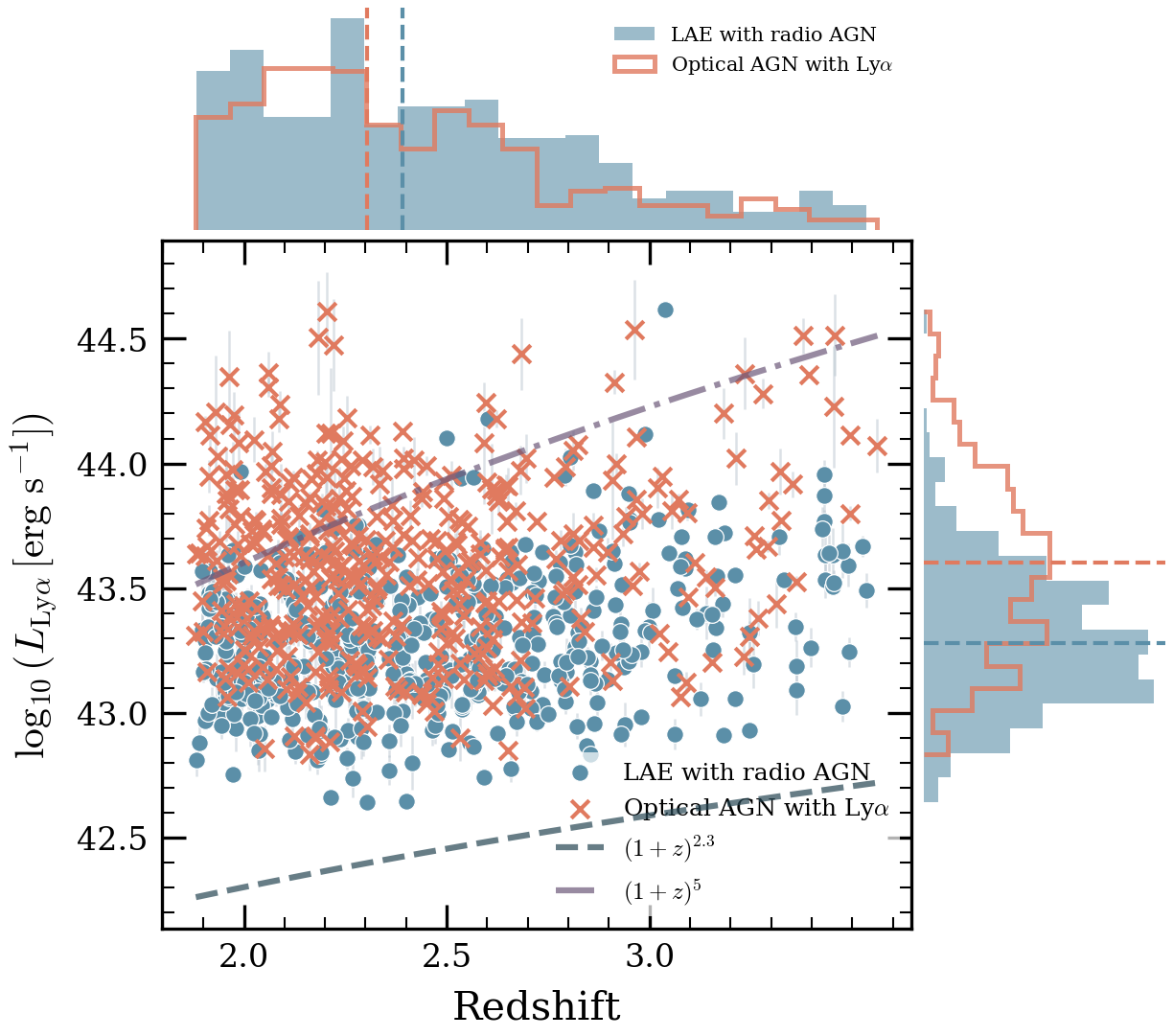}
        \caption{}
        \label{fig:Llya_z}
    \end{subfigure}

    \caption{Ly$\alpha$ and radio luminosities as a function of redshift  for `LAE with radio AGN' (blue circles and filled histograms) and `optical AGN with Ly$\alpha$ emission' (red crosses and step histograms). Vertical dashed lines in the histograms indicate the median values for each population.
    (a) Radio luminosity at 150 MHz versus redshift. The black dashed line indicates the LOFAR DR1 detection limit (5$\sigma$). Both populations span similar ranges in radio luminosity, with `optical AGN with Ly$\alpha$ emission' extending to slightly higher luminosities.
    (b) Ly$\alpha$ luminosity versus redshift. Purple and blue curves indicate $(1+z)^5$ and $(1+z)^{2.3}$ relations from \citet{zirm2009ApJ...694L..31Z}. `Optical AGN with Ly$\alpha$ emission' have a larger median Ly$\alpha$ luminosity than `LAE with radio AGN'.}

    \label{fig:luminosity_relations}
\end{figure*}

\subsubsection{Radio luminosities as a function of redshift}

Fig.~\ref{fig:luminosity_relations}\subref{fig:Lradio_z} shows the 150\,MHz radio luminosities of the `LAE with radio AGN' and `optical AGN with Ly$\alpha$ emission' samples as a function of redshift over $1.88 < z < 3.52$. The two populations have similar radio luminosity distributions, with median values of 
$\log_{10}(L_{\rm 150\,MHz}/\mathrm{W\,Hz^{-1}}) = 25.41\pm0.02$ and $25.44\pm0.02$, respectively. The `LAE with radio AGN' have a median redshift of $z = 2.39\pm{0.02}$, while the `optical AGN with Ly$\alpha$ emission' have $z = 2.30\pm{0.02}$. 
The luminosities range from $10^{24.6}$ to 
$10^{28.6}$\,W\,Hz$^{-1}$, placing our sources at the fainter end of the high-redshift radio galaxy (HzRG) population compared to \citet{Saxena2019MNRAS.489.5053S}, who studied HzRGs at $2 < z < 5.72$ with 150\,MHz luminosities of $10^{26.2}$--$10^{30}$\,W\,Hz$^{-1}$.

Our sources are above the AGN-dominated threshold of 
$L_{\rm 150\,MHz} > 10^{23.9}$\,W\,Hz$^{-1}$ 
(corresponding to $L_{\rm 1.4\,GHz} > 10^{23.2}$\,W\,Hz$^{-1}$ 
assuming $\alpha = 0.7$) identified by 
\citet{Calhau2020MNRAS.493.3341C}, consistent with our 
classification that both populations are AGN.

\begin{figure}
    \centering
        \includegraphics[width=\linewidth]{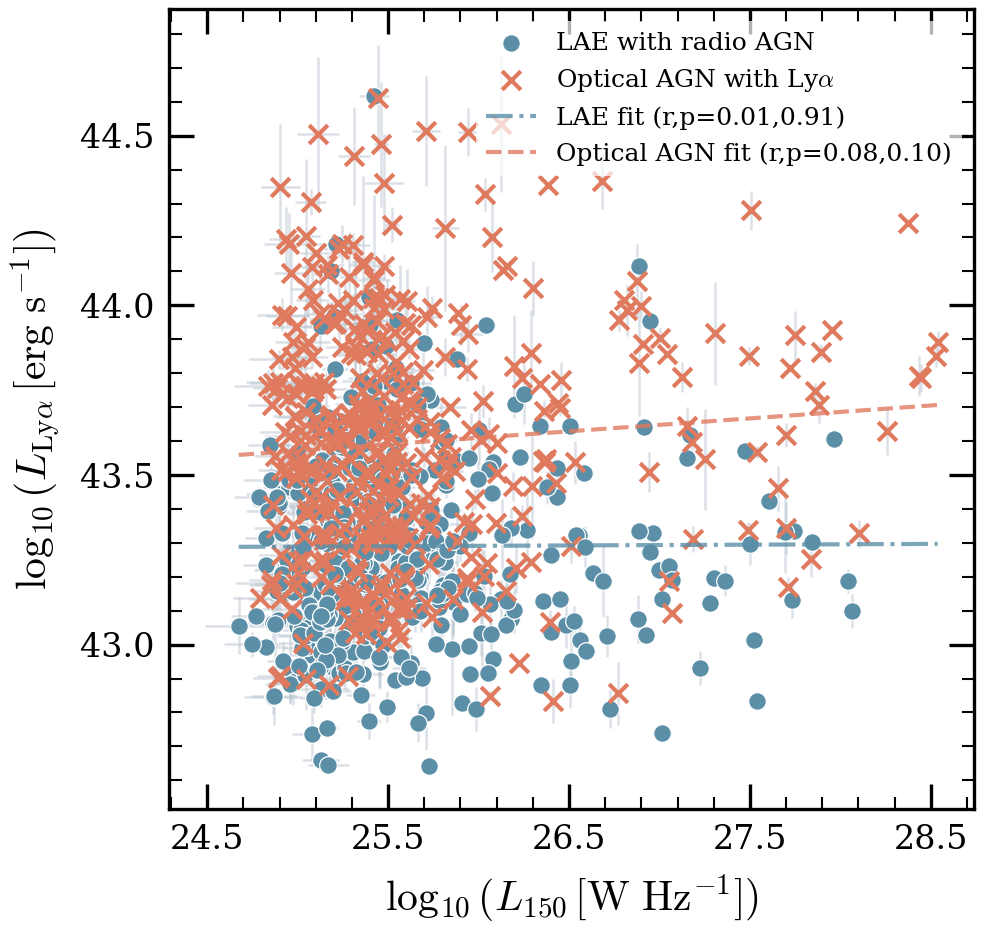}
        \caption{Ly$\alpha$ luminosity versus radio luminosity (log-scale). There is a weak negative correlation between Ly$\alpha$ and radio luminosity for `LAE with radio AGN' and a weak positive correlation for `optical AGN with Ly$\alpha$ emission'. Dashed lines show linear fits to each population. The lack of strong correlation suggests that Ly$\alpha$ and radio emission trace distinct physical processes or timescales in these systems.}
        \label{fig:Llya_Lradio}
\end{figure}

\subsubsection{Ly$\alpha$ luminosities as a function of redshift}

Fig.~\ref{fig:luminosity_relations}\subref{fig:Llya_z} shows the Ly$\alpha$ luminosity as a function of redshift for both samples. The `optical AGN with Ly$\alpha$ emission' exhibit systematically higher Ly$\alpha$ luminosities than the `LAE with radio AGN', with median values of  $\log_{10}(L_{\rm Ly\alpha}/\mathrm{erg\,s^{-1}}) = 43.60\pm0.02$ and $43.28\pm0.02$, respectively. This difference may partly reflect the different selection functions of the two samples: the `LAE with radio AGN' are first selected through Ly$\alpha$ emission and then classified using radio information, whereas the `optical AGN with Ly$\alpha$ emission' are first selected as optical AGN with detectable Ly$\alpha$ emission. The latter selection may preferentially include Ly$\alpha$-bright, AGN-dominated systems. This sample-level selection 
difference is consistent with our finding that the radio AGN fraction increases with Ly$\alpha$ luminosity (Section~\ref{fagn_lya}), 
and with the higher AGN fraction at the bright end of the 
Ly$\alpha$ luminosity function reported by \citet{Sobral2018MNRAS.477.2817S}, and likely contributes to the broader FWHM values seen in the optical AGN sample (Section~\ref{FWHM_z} and \ref{FWHM_L}).

Previous studies have reported mixed results on the evolution of Ly$\alpha$ luminosity in radio AGN. \citet{zirm2009ApJ...694L..31Z} found a positive evolution in HzRGs over $1 < z < 5$, with a dependence between $(1+z)^{2.3}$ and $(1+z)^{5}$, primarily driven 
by low-$L_{\rm Ly\alpha}$ sources at $z \sim 1$. In contrast, \citet{Saxena2019MNRAS.489.5053S} found no significant correlation for 13 faint HzRGs at $2 \lesssim z < 5.7$. 

For comparison, we overlay these scaling in 
Fig.~\ref{fig:Llya_z}. All our sources lie above the $(1+z)^{2.3}$ relation by $\sim 0.4$\,dex, and most fall between the two scaling. 
At $z \lesssim 2.5$, the `optical AGN with Ly$\alpha$ emission' tend to exceed the $(1+z)^{5}$ scaling, consistent with their higher Ly$\alpha$ luminosities powered by strong nuclear activity. 
Given the narrow redshift range ($1.88 < z < 3.52$) and flux-limited nature of our sample, we cannot robustly constrain the intrinsic evolution of Ly$\alpha$ luminosity. Larger samples extending to lower redshift will be needed to clarify this relation.

\subsubsection{Radio and Ly$\alpha$ luminosities}\label{bothL}

Fig.~\ref{fig:Llya_Lradio} shows the Ly$\alpha$ luminosity as a function of radio luminosity for both samples. Neither population shows a significant correlation: $r = 0.01$ ($p = 0.91$) for `LAE with radio AGN' and $r = 0.08$ ($p = 0.10$) for `optical AGN with Ly$\alpha$ emission'.

Our findings differ from the results reported by \cite{Jarvis2001MNRAS.326.1563J}, who found a positive correlation between Ly$\alpha$ and radio luminosities ($41.5 < \log_{10}(L_{\rm Ly\alpha}\ [\rm erg \ s^{-1}]) < 45.0$ and $27.0 < \log_{10}(L_{151}\ [\rm W \ Hz^{-1}]) < 28.5$) for sources at redshift $z>1.75$, and suggested that radio jets or quasar-driven shocks could photo-ionize the surrounding gas, enhancing Ly$\alpha$ emission.

However, our results are consistent with \cite{Saxena2019MNRAS.489.5053S}, who found no correlation between these two luminosities for faint HzRGs. Similarly, \citet{Calhau2020MNRAS.493.3341C} reported a flat relation for radio-detected LAEs, which they attributed to the different physical and temporal scales of radio and Ly$\alpha$ emission, with radio activity being comparatively long-lived. 

Although no correlation is seen between $L_{\rm Ly\alpha}$ and $L_{\rm 150\,MHz}$ for individual sources, the fraction of Ly$\alpha$-emitting sources increases with radio luminosity (Fig.~\ref{fig:f_L150}), and conversely, the radio detection fraction increases with Ly$\alpha$ luminosity (Fig.~\ref{fig:f_Llya}). This suggests that radio and Ly$\alpha$ emission are not directly linked, but both become more common in more powerful AGN.

Our results support the view that, at least for moderate-to-low radio luminosity sources, radio jet power and Ly$\alpha$ output are not directly coupled. Instead, Ly$\alpha$ emission appears to be more closely linked to the availability of ionized gas and the radiative output of the central engine rather than to the mechanical power of radio jets \citep[e.g.,][]{Heckman1991ApJ...370...78H}. For `optical AGN with Ly$\alpha$ emission', jet power and radiative activity may evolve independently \citep{Best2012MNRAS.421.1569B}, whereas in `LAEs with radio AGN' the observed Ly$\alpha$ emission is more likely driven by star formation or cooling radiation from the surrounding gas \citep{Smith2007MNRAS.378L..49S,Geach2009ApJ...700....1G, Dijkstra_2014}.

However, the sample spans a relatively narrow dynamic range in both Ly$\alpha$ and radio luminosity, and most sources lie close to the flux limits of HETDEX and LOFAR. This combination likely couples luminosity and redshift and restricts the accessible region of the parameter space. 
Deeper radio data or a broader luminosity baseline would allow a more robust assessment of whether the flat relation is physical or a product of sensitivity limits. 

\subsubsection{Radio spectral indices and Ly$\alpha$ luminosities}

\begin{figure}
    \centering
    \includegraphics[width=\linewidth]{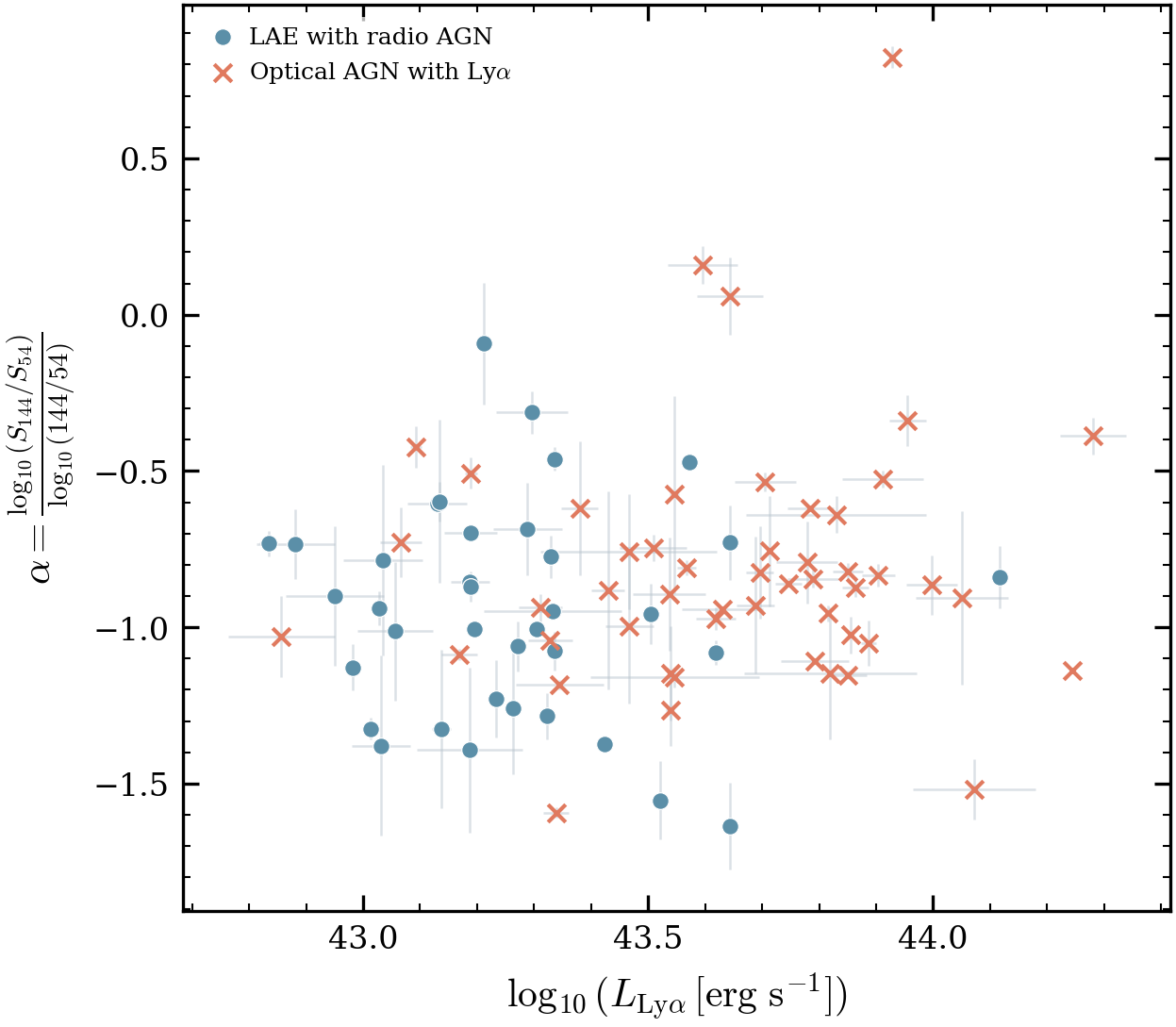}
        \caption{The low-frequency radio spectral index ($\alpha$) between 54 MHz and 144 MHz as a function of Ly$\alpha$ luminosity. Blue circles represent `LAEs with radio AGN', while red crosses show `optical AGN with Ly$\alpha$ emission'. No statistically significant correlation is observed between $\alpha$ and Ly$\alpha$ luminosity for either population, consistent with radio properties being independent of Ly$\alpha$ luminosity.} 
    \label{fig:alpha_Llya}
\end{figure}

We investigate the relationship between the low-frequency radio spectral index and Ly$\alpha$ luminosity for `LAE with radio AGN' and `optical AGN with Ly$\alpha$ emission'. The spectral index $\alpha$ is defined as:
\begin{equation}
    \alpha = \frac{\log_{10}(S_{144}/S_{54})}{\log_{10}(144/54)},
\end{equation}
where $S_{144}$ and $S_{54}$ are the flux densities at 144\,MHz from LoTSS and 54\,MHz from  the LOFAR LBA Sky Survey (LoLSS, \citealt{deGasperin2023A&A...673A.165D}), respectively.

Figure~\ref{fig:alpha_Llya} presents the spectral index as a function of Ly$\alpha$ luminosity for both source populations. The `LAE with radio AGN' exhibit a median spectral index of $\alpha = -0.94\pm0.07$, while the `optical AGN with Ly$\alpha$ emission' show a median of $\alpha =-0.87\pm0.05$. For both samples, the median spectral indices are slightly steeper than the typical radio selected AGN values of $-0.73$ \citep{Smol2017A&A...602A...1S}. However, the steep spectral indices confirm that our sources are powered by AGN activity rather than star formation. We find no statistically significant correlation between the 
radio spectral index and Ly$\alpha$ luminosity for either 
population: a Spearman rank  test yields $r = 
-0.08$ ($p = 0.63$) for `LAE with radio AGN' and $r = 0.09$ ($p = 0.51$) for `optical AGN with Ly$\alpha$ emission'. This is consistent 
with the findings of \citet{Calhau2020MNRAS.493.3341C}, who similarly reported no significant relation between the 1.4--3.0\,GHz spectral index and Ly$\alpha$ luminosity for LAEs at $z \sim 2$--6 in the SC4K survey.

The absence of a correlation suggests that the radio spectral properties and Ly$\alpha$ emission trace physically distinct processes or operate on different timescales. The low-frequency radio spectral index reflects the age of the synchrotron-emitting electron in the jets \citep{Morganti2017FrASS...4...42M}, while Ly$\alpha$ luminosity is more sensitive to instantaneous AGN accretion activity or star formation \citep{Hickox2014ApJ...782....9H}. Taking together with the absence of a correlation between radio and Ly$\alpha$ luminosities discussed in Section \ref{bothL}, this supports a picture in which the radio and Ly$\alpha$ emission are largely decoupled.

\begin{figure*}
    \centering
    % --- (a) ---
    \begin{subfigure}{0.5\textwidth}
        \centering
        \includegraphics[width=\linewidth]{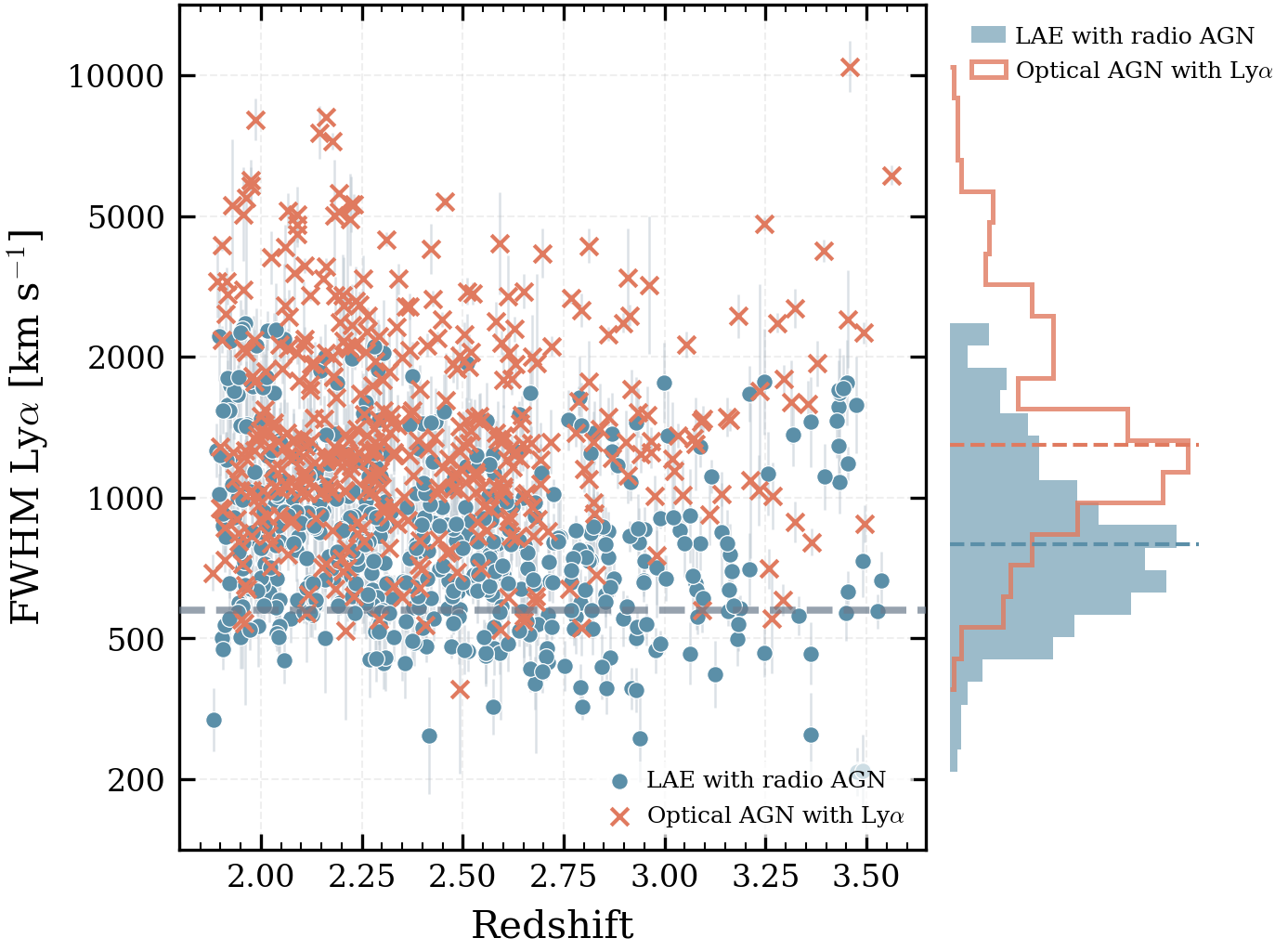}
        \caption{}
        \label{fig:fwhm-z}
    \end{subfigure}
    \hfill
    % --- (b) ---
    \begin{subfigure}{0.45\textwidth}
        \centering
        \includegraphics[width=\linewidth]{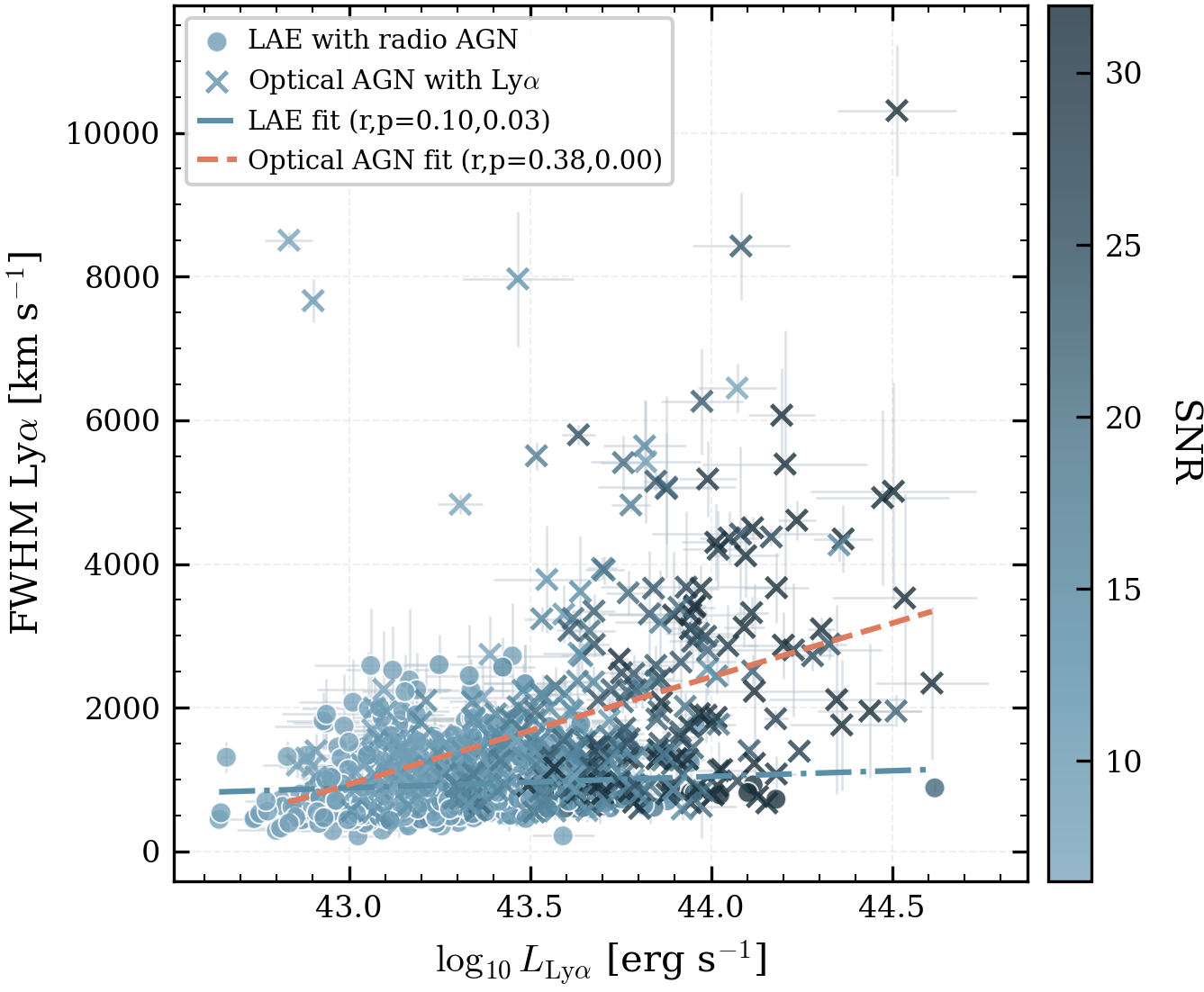}
        \caption{}
        \label{fig:fwhm-Llya}
    \end{subfigure}

    \caption{Correlations between FWHM and redshift or Ly$\alpha$ luminosity for `LAE with radio AGN' (circles) and `optical AGNs with Ly$\alpha$ emission' (crosses). Blue dash dot and red dashed lines represent the best linear fits for each population. 
    (a) FWHM versus redshift. The marginal histogram (right panel) illustrates the FWHM distribution for both samples; dashed lines  respective median values. The non-uniform y-axis scaling emphasizes the large spread in line widths while maintaining visibility across the full dynamic range. The gray line marks FWHM = 600 $\rm km \ s^{-1}$.
    (b) FWHM versus Ly$\alpha$ luminosity.  Both samples show positive correlations. The colour-bar represents the SNR of Ly$\alpha$ line, 
showing no systematic trend with FWHM within fixed luminosity bins.}
    \label{fig:fwhm}
\end{figure*}

\subsection{FWHM}

\subsubsection{FWHM and redshift} \label{FWHM_z}

Fig.~\ref{fig:fwhm-z} shows the FWHM of the Ly$\alpha$ emission line as a function of redshift for both samples. The `LAE with radio AGN' exhibit systematically lower FWHM values than the `optical AGN with Ly$\alpha$ emission', with median values of $833\pm17$\,km\,s$^{-1}$ and 
$1374\pm38$\,km\,s$^{-1}$, respectively. This difference is expected given the selection criteria: the `optical AGN with Ly$\alpha$ emission' in HETDEX are classified partly through the presence of broad lines with FWHM $> 1000$\,km\,s$^{-1}$ \citep{Liu2022ApJ...935..132L}, which preferentially selects sources with larger FWHMs.

The `LAE with radio AGN' show a weak but statistically significant negative correlation between FWHM and redshift ($r = -0.18$, $p < 0.001$), while no significant trend is seen for the `optical AGN with Ly$\alpha$ emission' ($r = -0.03$, $p = 0.56$). 
The negative trend for `LAE with radio AGN' is qualitatively consistent with \citet{Saxena2019MNRAS.489.5053S}, who reported a similar negative correlation for faint HzRGs, and interpreted it as evidence that the gas kinematics or physical conditions of Ly$\alpha$-emitting regions evolve with redshift.

\subsubsection{FWHM and Ly$\alpha$ luminosities}\label{FWHM_L}

Fig. \ref{fig:fwhm-Llya} presents the relationship between FWHM and Ly$\alpha$ luminosity. We observe a modest positive correlation for `optical AGN with Ly$\alpha$ emission' ($r=0.38,p<0.01$) and a weak positive correlation for `LAE with radio AGN' ($r=0.10,p=0.03$). The `optical AGN with Ly$\alpha$ emission' exhibit larger scatter compared to `LAE with radio AGN'. Additionally, at high Ly$\alpha$ luminosities, `optical AGN with Ly$\alpha$ emission' typically show broader FWHM values than LAEs. 

This positive correlation can be naturally explained by the fact that both quantities trace the strength of the energy input into the ionized gas. Stronger ionizing sources, such as powerful AGN or intense starbursts, produce more Ly$\alpha$ photons while simultaneously driving more turbulent and dynamic gas kinematics, resulting in broader emission lines \citep[e.g.,][]{Heckman1991ApJ...370...78H,Cai2018ApJ...861L...3C,Saxena2019MNRAS.489.5053S}.

The `LAE with radio AGN' and `optical AGN with Ly$\alpha$ emission' have lower Ly$\alpha$ luminosities and wider FWHM values compared to the literature HzRGs in \cite{Saxena2019MNRAS.489.5053S}. In our work, only a small fraction of sources fall within the star-forming galaxy region and exhibit relatively narrow FWHM values, while the majority show broad FWHM profiles. \cite{Saito2008ApJ...675.1076S} find a positive relationship between these two parameters for 18 extended Ly$\alpha$ sources with EW $>100\AA$. \cite{Sobral2018MNRAS.477.2817S} also find that Ly$\alpha$ FWHM tends to increase with Ly$\alpha$ luminosity in luminous LAEs at similar redshift, with the trend mainly driven by AGN. At higher redshift, however, the comparison is less direct because the samples are primarily star-forming LAEs. \cite{Songaila2024ApJ...971..136S}, including comparisons with the samples of \cite{Ning2020ApJ...903....4N} and \cite{Shibuya2018PASJ...70S..15S}, found that high-redshift LAEs occupy broadly consistent ranges in Ly$\alpha$ luminosity and FWHM, rather than showing a clear positive FWHM--$L_{\rm Ly\alpha}$ correlation. Therefore, these high-redshift studies are used here mainly as a reference for the typical parameter space of star-forming LAEs, not as a direct comparison to our AGN-dominated LAE sample. The broader FWHM values in our sample may result from differences in sample selection, redshift, and the physical nature of the sources, especially the contribution of AGN activity to the Ly$\alpha$ line width.

We also examined potential flux-dependent detection biases by analysing the relationship between FWHM and SNR of Ly$\alpha$ line. At fixed integrated flux, broader lines have lower peak flux and thus lower SNR, potentially leading to preferential detection of narrow-line sources at low luminosities. Within fixed Ly$\alpha$ luminosity bins, sources span a range of fluxes due to varying redshift. We find no significant correlation between FWHM and SNR within these bins, suggesting that we don't miss broad-line sources at low fluxes. This indicates the observed FWHM-luminosity correlation is not primarily an artifact of flux-dependent selection effects.

Nevertheless, we note several instrumental and measurement limitations. The instrumental resolution of HETDEX (R $\sim$ 800) also limits the ability to measure very narrow Ly$\alpha$ lines. At the faint end, low SNR spectra may further broaden the measured profiles. Combined with the limited dynamic range of the sample (roughly 800–2000 $\rm km\ s^{-1}$), these effects could weaken any true underlying correlations in Section \ref{FWHM_z} and \ref{FWHM_L}.

\subsubsection{FWHM and size}\label{size}
\begin{figure}
    \centering
    \includegraphics[width=0.8\linewidth]{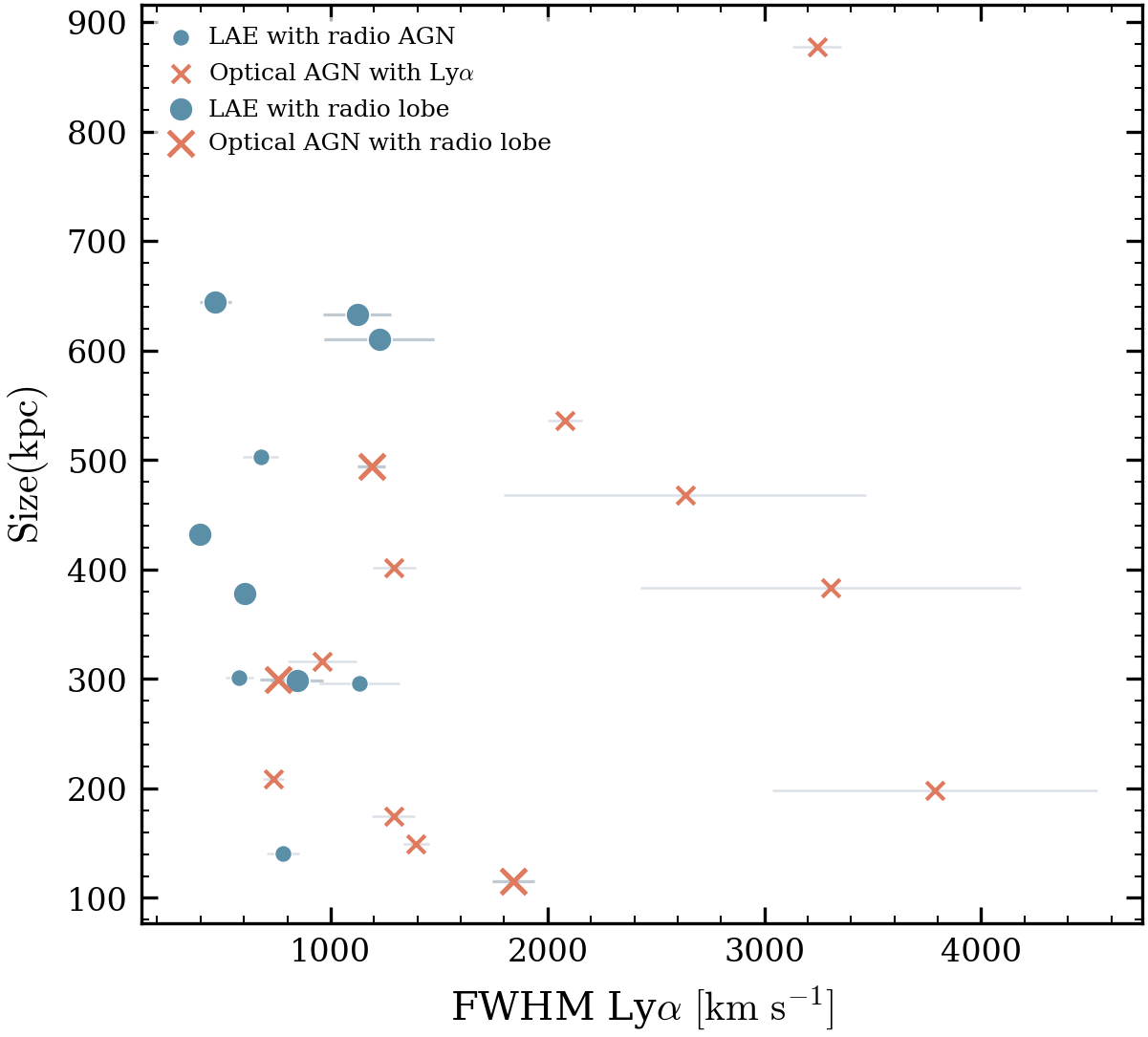}
        \caption{The radio size as a function of Ly$\alpha$ FWHM. Blue circles represent `LAEs with radio AGN', while red crosses show `optical AGN with Ly$\alpha$ emission'. Larger markers indicate sources with visually identified radio lobes. No statistically significant correlation is observed between radio size and Ly$\alpha$ FWHM for either population.} 
    \label{fig:size_fwhm}
\end{figure}

We investigate the relationship between FWHM and radio size for our sample. The size information was obtained from the LoTSS DR2 catalogue with optical counterparts \citep{Hardcastle2023A&A...678A.151H}. In our sample, there are 23 sources resolved, out of which 10 are `LAE with radio AGN' and 13 is `optical AGN with Ly$\alpha$ emission'. We summarize the basic information of these sources in Table \ref{tab:extended} in Appendix. 

Fig.~\ref{fig:size_fwhm} presents the radio size as a function of Ly$\alpha$ FWHM. For the `LAE with radio AGN', the sizes range from 140 to 645 kpc. The `optical AGN with Ly$\alpha$ emission' have size between 150 and 878 kpc. No statistically significant correlation is observed between radio size and Ly$\alpha$ FWHM for either population, likely due to the small sample size.

In a jet-gas interaction scenario, an anti-correlation between radio size and Ly$\alpha$ FWHM might be expected \citep{Saxena2019MNRAS.489.5053S}. When the radio source is compact, the jet is still propagating through the interstellar medium (ISM), transferring kinetic energy to the surrounding gas and broadening the emission line \citep{vanOjik1997A&A...317..358V}. As the radio source expands beyond the line-emitting region, the gas kinematics relax, resulting in narrower line widths. However, we find no significant correlation between these two parameters in our sample, suggesting that jet-gas interaction may not be the dominant mechanism driving Ly$\alpha$
 line broadening in these sources. Alternatively, the lack of correlation could reflect the diversity of jet orientations, host galaxy properties, or the limited dynamic range of our resolved sample.

 In Fig. \ref {fig:extended_sources}, among the 23 resolved radio sources in our sample, 9 exhibit extended radio lobes, clearly indicating AGN driven jet activity. The remaining 14 sources do not show clear lobes. These sources may host compact or young jets, their lobe emission may be too faint to detect with current sensitivity, or projection effects may obscure their morphology.

The relatively high incidence ($39\%=9/23$) of lobe-bearing sources in such a small sample highlights the importance of jet-driven mechanical feedback in a subset of our LAE and AGN population. This fraction should be considered a lower limit, as LoTSS may not fully resolve all lobe-dominated systems. Nevertheless, the presence of extended radio lobes in some sources may have significant implications for the kinematics of the Ly$\alpha$ emitting gas, especially in terms of line broadening or spatial extension.

\section{Discussion}\label{discussion}
\subsection{Increasing radio AGN fraction toward brighter Ly$\alpha$ emitting sources}

Previous studies have consistently shown a positive relationship between AGN fraction and Ly$\alpha$ luminosity. For instance, \citet{Sobral2018MNRAS.477.2817S} analyzed 21 spectroscopically-confirmed luminous LAEs at redshift $2<z<3$, finding an AGN fraction as high as $60\pm11\%$. In their sample, nearly all LAEs brighter than $10^{43.3}\,\rm erg \ s^{-1}$ host an AGN. They also highlighted earlier studies which reported similarly high AGN fractions (20–50\%) among luminous LAEs at lower redshift ($z=0.3$–$1$ and $z=2$–$3$), contrasted with much lower AGN fractions (just a few percent) among fainter, more numerous LAEs. They explored the relationship between FWHM and Ly$\alpha$ luminosity, noting only a weak or negligible correlation in broad-line AGNs, possibly explained by varying the Ly$\alpha$ escape fraction.

Similarly, \citet{Calhau2020MNRAS.493.3341C} examined around 4000 LAEs from the SC4K survey at $2.2<z<6$ in the COSMOS field. They found an overall AGN fraction of 8.6\% (X-ray and radio combined), with clear trends of increasing AGN fractions at higher Ly$\alpha$ luminosities. They also observed that AGN fraction decreases with increasing redshift. Specifically, radio AGN comprised only 3.2\% of the sample, lower than the X-ray AGN fraction, which reaches about 20–30\% at higher luminosities. Their findings align closely with previous results from \citet{Matthee2017MNRAS.471..629M} and \citet{Wold2014ApJ...783..119W, Wold2017ApJ...848..108W}.  \citet{Matthee2017MNRAS.471..629M} studied around 1000 emission-line galaxies in the Boötes field, detecting 8 X-ray sources out of 41 LAEs at $z=2.2$. They concluded that bright LAEs, predominantly AGN, strongly influence the luminous end of the Ly$\alpha$ luminosity function. Likewise, studies by \citet{Wold2014ApJ...783..119W,Wold2017ApJ...848..108W}, utilizing GALEX data at lower redshift ($0.3<z<2.2$), further confirmed this positive correlation between AGN fraction and Ly$\alpha$ luminosity ($41.5 < \rm log_{10}(L_{\rm Ly\alpha, erg \ s^{-1}})<43.5$).  These results reinforce the idea that luminous LAEs are significantly more likely to host AGNs, indicating an increasing role of nuclear ionizing radiation at the luminous end. In addition, within the `optical AGN with Ly$\alpha$ emission' and `LAE with radio AGN' subsamples, Ly$\alpha$ luminosities show little redshift evolution within $1.88<z<3.52$. This is broadly consistent with previous studies showing that the global LAE population and Ly$\alpha$ luminosity functions evolve only mildly over $z\sim2$--6 \citep{2020ARA&A..58..617O}. This may suggest that, over this redshift range, the observed Ly$\alpha$ luminosity is strongly affected by local radiative-transfer and gas conditions, such as dust, geometry, and IGM attenuation, in addition to the intrinsic ionizing output of the central source. 

Our results are subject to several caveats. First, the limited sample size and the heterogeneous redshift coverage restrict the statistical significance of the derived AGN fractions. Second, AGN classifications based on optical or radio diagnostics are incomplete, as obscured AGN or compact radio sources may be missed. Third, Ly$\alpha$ luminosity is shaped not only by ionizing radiation but also by complex radiative transfer effects, making it an indirect tracer of AGN activity. Taken together, our results support a picture in which most LAEs are powered by star formation, with AGN activity contributing only in a small subset of luminous or radio-detected sources. This picture highlights the diversity of mechanisms driving Ly$\alpha$ emission and emphasizes the need for multi-wavelength diagnostics to disentangle their origins.

\subsection{Fraction of radio sources with Ly$\alpha$ emission}\label{f2}

\begin{figure}
    \centering
    \includegraphics[width=0.8\linewidth]{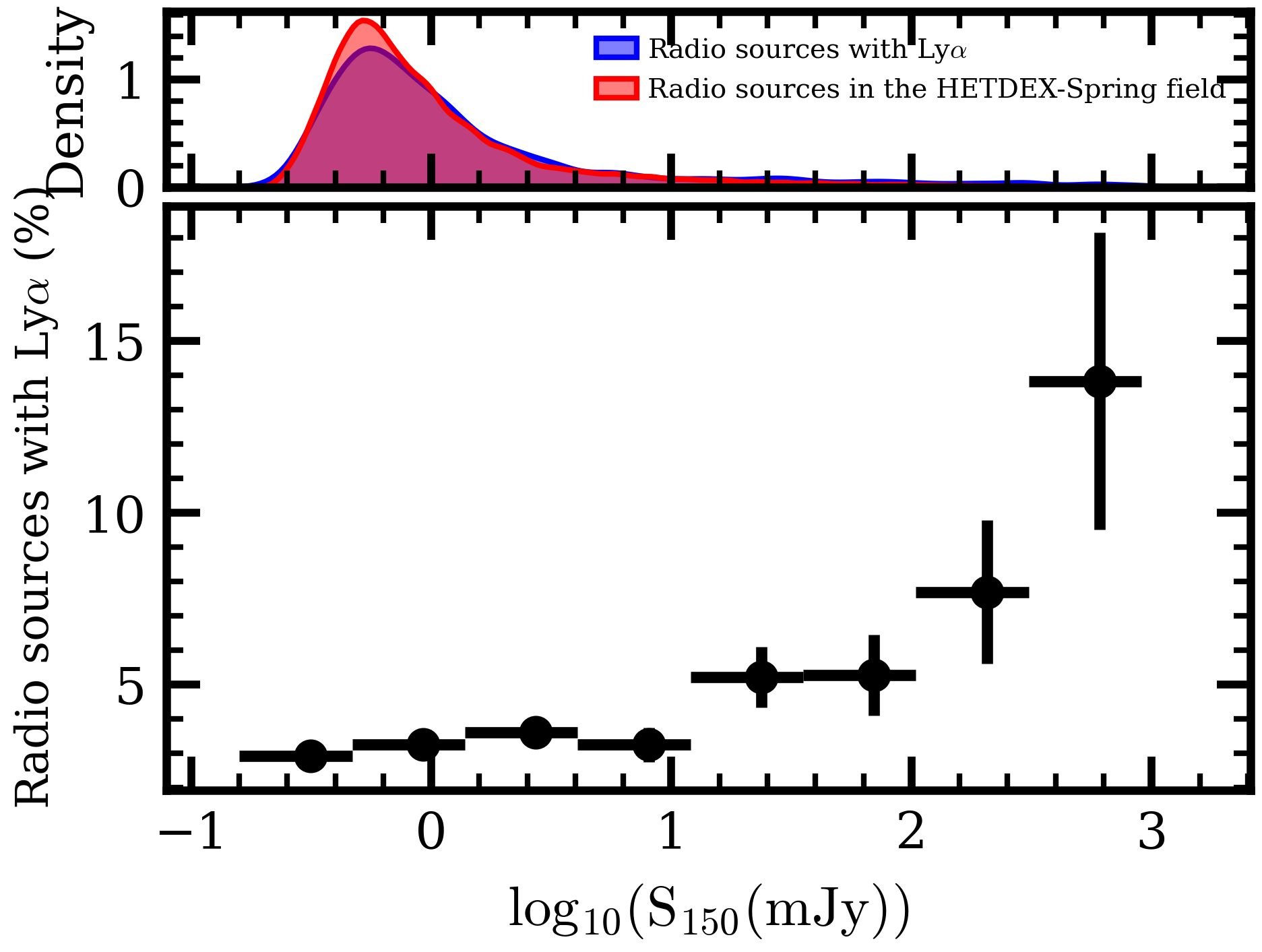}
    \includegraphics[width=0.8\linewidth]{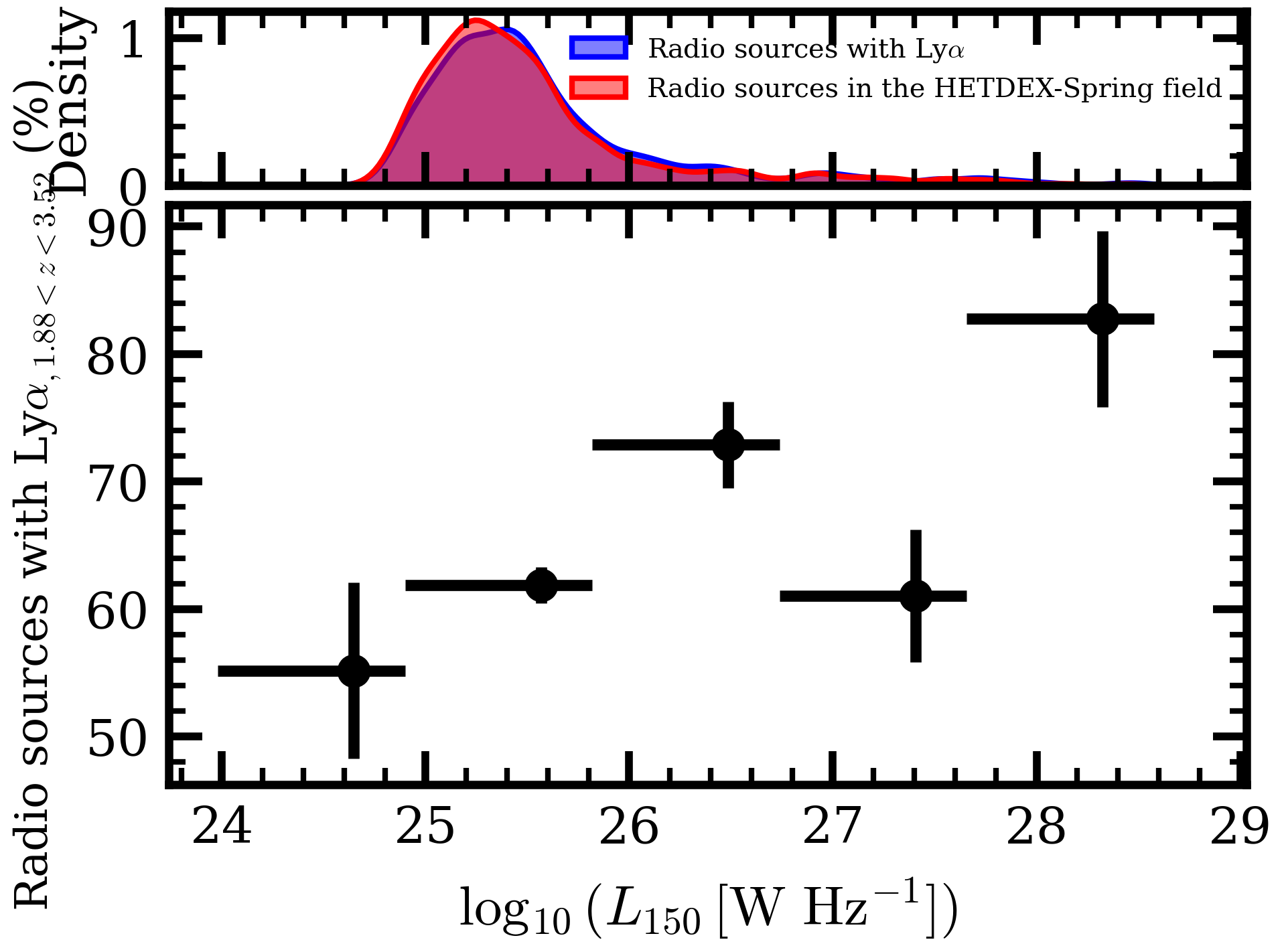}
    \caption{\textit{Top}: Fraction of radio sources with Ly$\alpha$ emission vs 150 MHz flux densities. \textit{Bottom:} The same fraction as a function of 150 MHz radio luminosities within $1.88<z<3.52$. Both panels show a positive correlation.}
    \label{fig:s144_fraction}
\end{figure}

A total of  \flaec \ of the radio sources in our sample exhibit Ly$\alpha$ emission. We also found a constant trend between this fraction and the 150 MHz radio luminosity at the intermediate radio luminosity bins. However, since the calculation of radio luminosity requires redshift information, we suspect that this trend may be a consequence of redshift-dependent selection biases in the sample construction. To further investigate this effect, we analyzed this fraction as a function of 150 MHz flux density (in the top panel of Fig. \ref{fig:s144_fraction}). The positive correlation between flux density and $f_{\rm Ly\alpha}$ suggests that the observed flat trend between luminosity is indeed a consequence of redshift bias. As a quantitative check, we recompute the same fraction as a function of 150 MHz radio luminosities, restricted to sources within $1.88<z<3.52$. As shown in the bottom panel of Fig. \ref{fig:s144_fraction}, this restricted sample also show a positive correlation between these two parameters. This supports the interpretation that the flat trend seen in the full luminosity-based analysis is partly driven by redshift-dependent selection effects.

In addition, the limited dynamic range and the mixture of AGN-dominated and star-forming radio sources in the lower-luminosity bins may dilute intrinsic correlations. Future studies with larger samples spanning more uniform redshift distributions would be needed to better constrain intrinsic physical trends and separate them from the effects of flux limits and sample heterogeneity.

To place our finding in a broader context, we compare our results with studies of $f_{\rm Ly\alpha}$ among other galaxy populations. Studies by \citet{Stark2010MNRAS.408.1628S,Stark2011ApJ...728L...2S} and \citet{Schenker2012ApJ...744..179S} used Keck spectroscopy to investigate the fraction of Ly$\alpha$ emission among Lyman break galaxies (LBGs) across redshift $z=3-8$. They observed an increase in the fraction of galaxies exhibiting Ly$\alpha$ emission from $z=3 \ (\sim10\%)$ to $z=6 \ (\sim20\%)$, followed by a decrease at higher redshift ($z=6-8$, from $\sim15\%$ to $\sim10\%$). Additionally, they reported a positive correlation between the fraction of galaxies showing Ly$\alpha$ emission and UV luminosity.  

Our investigation extends these results to the population of radio-selected sources. The increased $f_{\rm Ly\alpha}$ among radio bright sources might reflect an underlying connection between Ly$\alpha$ emission and AGN activity, as suggested by previous studies (e.g., \citealt{Overzier2013ApJ...771...89O,Cai2017ApJ...837...71C}). AGNs contribute to Ly$\alpha$ emission through both ionizing radiation and outflow-drive shock ionization \citep{Geach2009ApJ...700....1G,Dijkstra_2014}, but the absence of a correlation between radio and Ly$\alpha$ luminosities indicates that once Ly$\alpha$ emission is triggered, its strength is governed primarily by local gas conditions and radiative transfer rather than by the absolute jet power.

\section{Conclusions}\label{conclusion}

Our results provide several insights into the interplay between radio activity and Ly$\alpha$ emission at $2<z<4$:

\begin{enumerate}
  \item Radio AGN fraction of Ly$\alpha$-emitting sources. Only a small fraction (\fagn) of Ly$\alpha$-emitting sources host radio AGN; this fraction increases to \fagnc\ after completeness corrections. This measurement suggests that, although AGN can power Ly$\alpha$ emission, most Ly$\alpha$-emitting sources at these redshift are likely powered by star formation. However, the radio AGN fraction increases with Ly$\alpha$ luminosity, indicating that brighter Ly$\alpha$ sources are more prone to hosting significant black hole activity. This implies that more massive or actively growing galaxies can sustain both intense star formation and AGN activity simultaneously.

  \item Ly$\alpha$ fraction of radio sources. Only a small fraction (\flae) of radio sources exhibit Ly$\alpha$ emission, increasing significantly to \flaec\ after completeness corrections. This fraction exhibits a positive correlation with radio 
luminosity, although the trend flattens at intermediate 
luminosities before resuming growth at higher radio powers. A possible scenario is that strong ionizing radiation from an AGN can contribute to enhanced Ly$\alpha$ emission.

  \item Radio luminosity vs. redshift. Both populations show similar radio luminosity distributions above the AGN threshold 
($L_{\rm 150\,MHz} > 10^{23.9}$\,W\,Hz$^{-1}$), confirming 
AGN-dominated radio emission.

  \item Ly$\alpha$ luminosity vs. redshift. `Optical AGN with Ly$\alpha$ emission' sources are systematically more luminous in Ly$\alpha$ than  `LAE with radio AGN', reflecting their different selection methods.

  \item Ly$\alpha$ vs. radio luminosity. No significant correlation exists between Ly$\alpha$ and radio luminosity, which suggests that the mechanisms responsible for Ly$\alpha$ emission in these systems are largely decoupled from those driving radio output.

  \item Low-frequency spectral index ($\alpha$) vs. radio luminosity. No significant correlation between $\alpha$ and radio luminosity further emphasizes that spectral properties of the radio emission do not directly trace the Ly$\alpha$ emission mechanisms.

  \item FWHM trends. Ly$\alpha$ line FWHM shows no 
significant correlation with redshift, suggesting that line widths 
do not systematically evolve over $1.88 < z < 3.52$, despite potentially stronger feedback at higher redshift. However, FWHM correlates positively with Ly$\alpha$ luminosity, supporting the hypothesis that more luminous systems exhibit broader lines due to enhanced velocity dispersions or more powerful outflows. We identify 23 
spatially resolved radio sources with Ly$\alpha$ emission (projected 
sizes 150–878 kpc), 39\% of which display lobe-like morphologies 
indicative of AGN-driven radio jets. Notably, we find no correlation between FWHM and radio source size, indicating 
minimal direct jet-gas interaction in these systems.
\end{enumerate}

Overall, these findings support a picture in which most Ly$\alpha$-emitting sources are powered predominantly by star formation rather than AGN activity. However, AGN activity can trigger or enhance Ly$\alpha$ emission, as evidenced by the increasing AGN fraction with Ly$\alpha$ luminosity and the rising Ly$\alpha$ fraction with radio luminosity. The lack of correlations between Ly$\alpha$ luminosity and both radio luminosity and low-frequency spectral index, as well as between FWHM and radio size, reveals that Ly$\alpha$ and radio emission are decoupled. This indicates that the strength of Ly$\alpha$ emission is regulated by additional factors, such as gas content, geometry, and radiative transfer, rather than being directly scaled with jet power.

Future observations, particularly those offering deeper radio coverage or better spatial resolution for both radio and optical data, will be crucial for connecting these broad statistical trends to the intricate details of feedback processes, gas dynamics, and the life cycle of high-redshift galaxies with Ly$\alpha$ emission.

\begin{acknowledgements}
LOFAR data products were provided by the LOFAR Surveys Key Science project (LSKSP; https://lofar-surveys.org/) and were derived from observations with the International LOFAR Telescope (ILT). LOFAR \citep{vanHaarlem2013A&A...556A...2V} is the Low Frequency Array designed and constructed by ASTRON. It has observing, data processing, and data storage facilities in several countries, which are owned by various parties (each with their own funding sources), and which are collectively operated by the ILT foundation under a joint scientific policy. The efforts of the LSKSP have benefited from funding from the European Research Council, NOVA, NWO, CNRS-INSU, the SURF Co-operative, the UK Science and Technology Funding Council and the Jülich Supercomputing Centre. 
\newline
      HETDEX is led by the University of Texas at Austin McDonald Observatory and Department of Astronomy with participation from the Ludwig-Maximilians-Universität München, Max-Planck-Institut für Extraterrestrische Physik (MPE), Leibniz-Institut für Astrophysik Potsdam (AIP), Texas A\&M University, Pennsylvania State University, Institut für Astrophysik Göttingen, The University of Oxford, Max-Planck-Institut für Astrophysik (MPA), The University of Tokyo and Missouri University of Science and Technology.

        Observations for HETDEX were obtained with the Hobby-Eberly Telescope (HET), which is a joint project of the University of Texas at Austin, the Pennsylvania State University, Ludwig-Maximilians-Universität München, and Georg-August-Universität Göttingen. The HET is named in honor of its principal benefactors, William P. Hobby and Robert E. Eberly. The Visible Integral-field Replicable Unit Spectrograph (VIRUS) was used for HETDEX observations. VIRUS is a joint project of the University of Texas at Austin, Leibniz-Institut für Astrophysik Potsdam (AIP), Texas A\&M University, Max-Planck-Institut fürExtraterrestrische Physik (MPE), Ludwig-Maximilians-Universität München, Pennsylvania State University, Institut für Astrophysik Göttingen, University of Oxford, and the Max-Planck-Institut fur Astrophysik (MPA).

        The authors acknowledge the Texas Advanced Computing Center (TACC) at The University of Texas at Austin for providing high performance computing, visualization, and storage resources that have contributed to the research results reported within this paper. URL: http://www.tacc.utexas.edu.

        Funding for HETDEX has been provided by the partner institutions, the National Science Foundation, the State of Texas, the US Air Force, and by generous support from private individuals and foundations.
    LKM is grateful for support from a UKRI FLF [MR/Y020405/1]. 
    C.G. acknowledges support from the National Science Foundation under grant AST-2408358. 
\end{acknowledgements}

\bibliographystyle{aa} 
\bibliography{aa}

\begin{appendix}
\onecolumn

\section{Additional Figure}

\begin{figure*}[!ht]
  \centering
  \setlength{\abovecaptionskip}{2pt}
  \setlength{\belowcaptionskip}{2pt}
  
  %---- Row 1 ----%
  \includegraphics[width=0.188\textwidth]{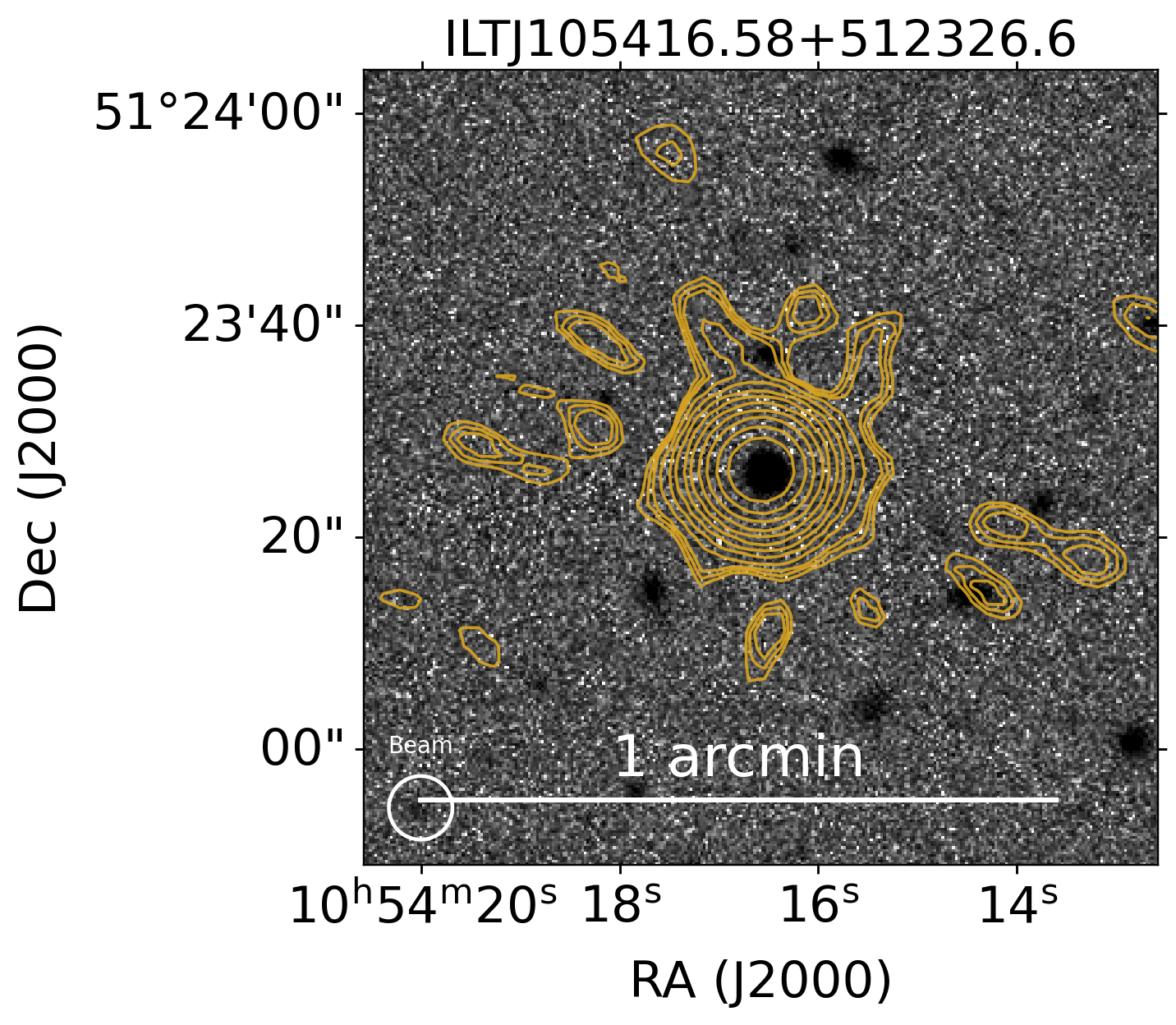}\hfill
  \includegraphics[width=0.188\textwidth]{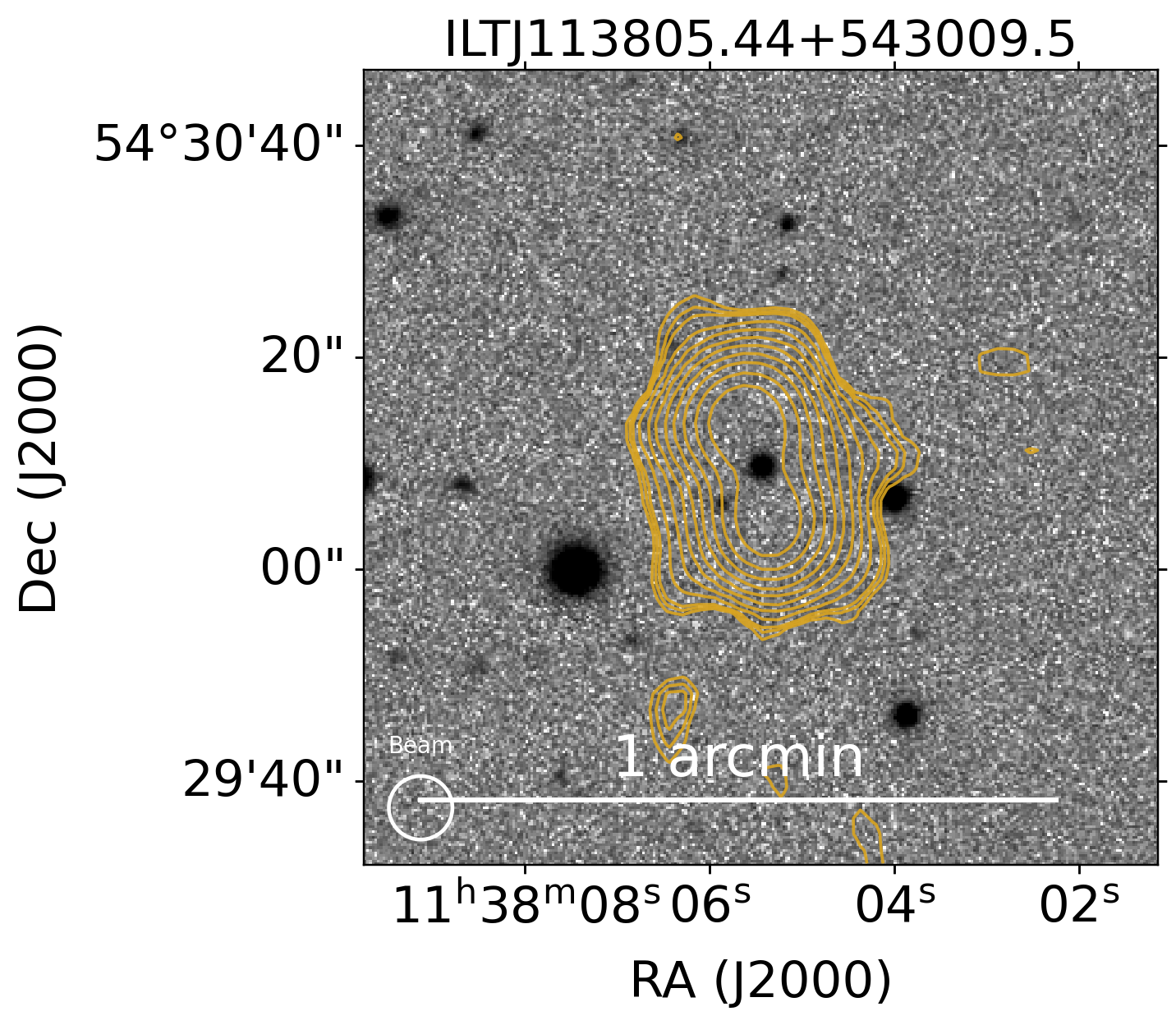}\hfill
  \includegraphics[width=0.188\textwidth]{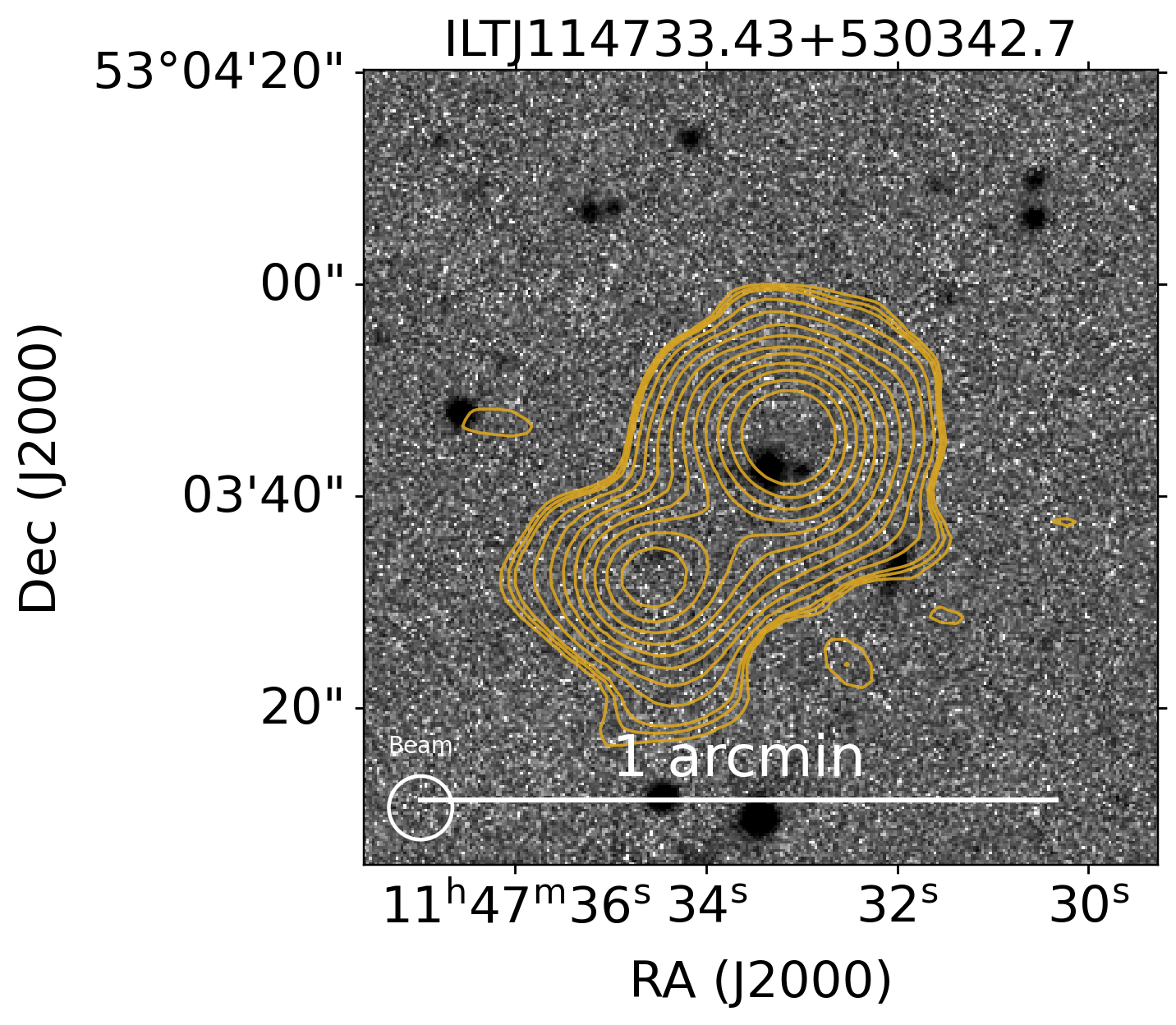}\hfill
  \includegraphics[width=0.188\textwidth]{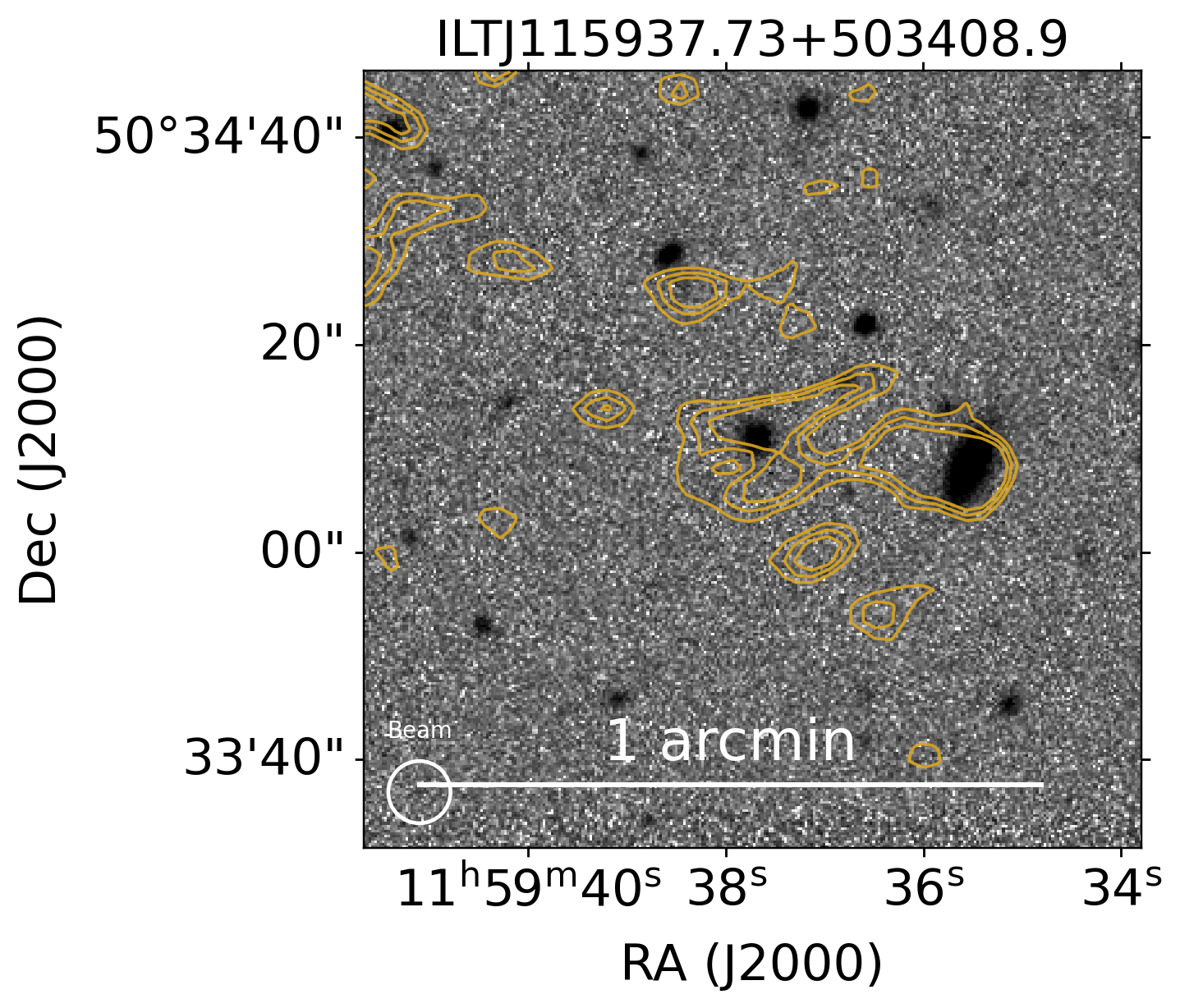}\hfill
  \includegraphics[width=0.188\textwidth]{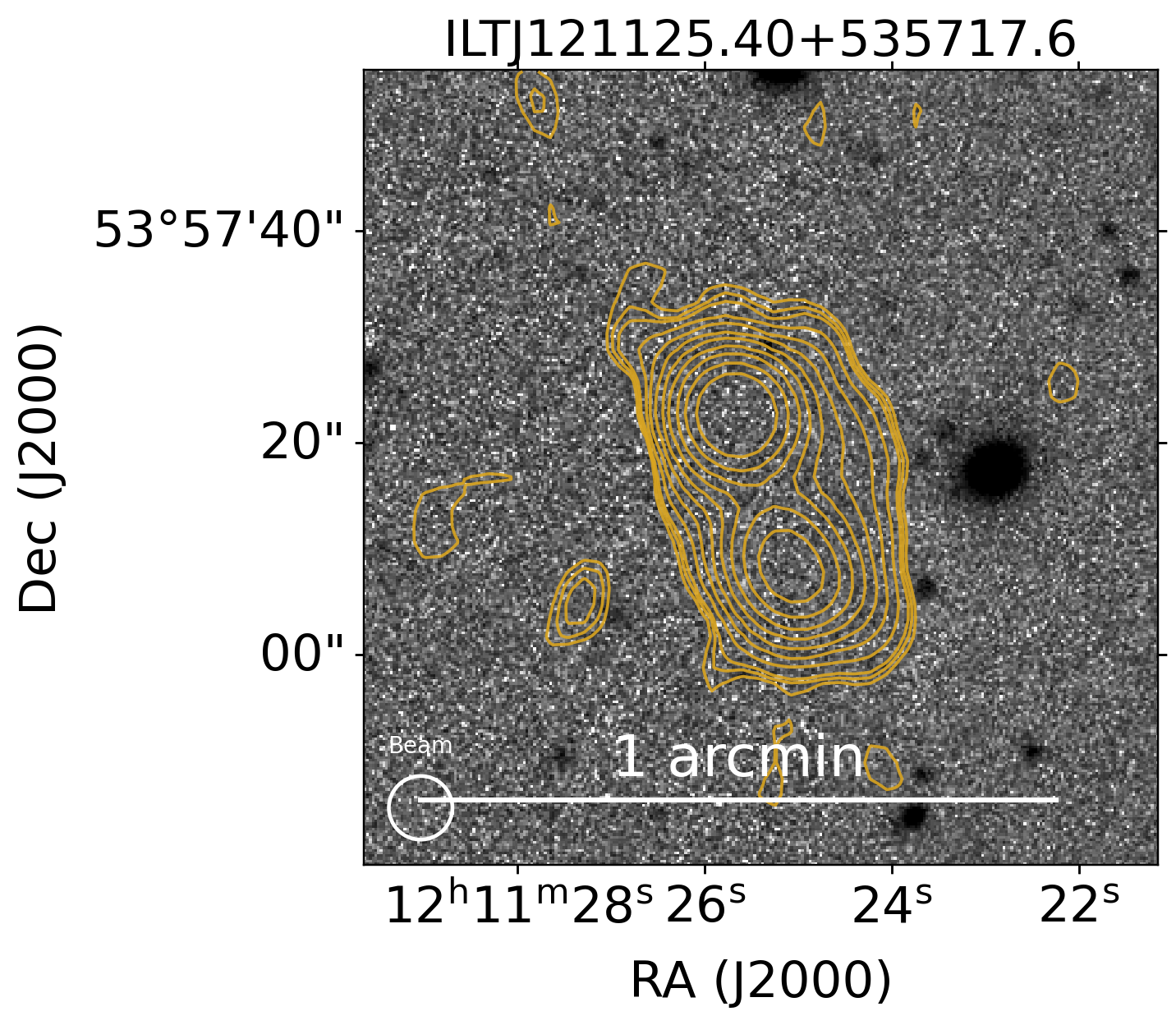}
  
  \par\vspace{1pt}
  
  %---- Row 2 ----%
  \includegraphics[width=0.188\textwidth]{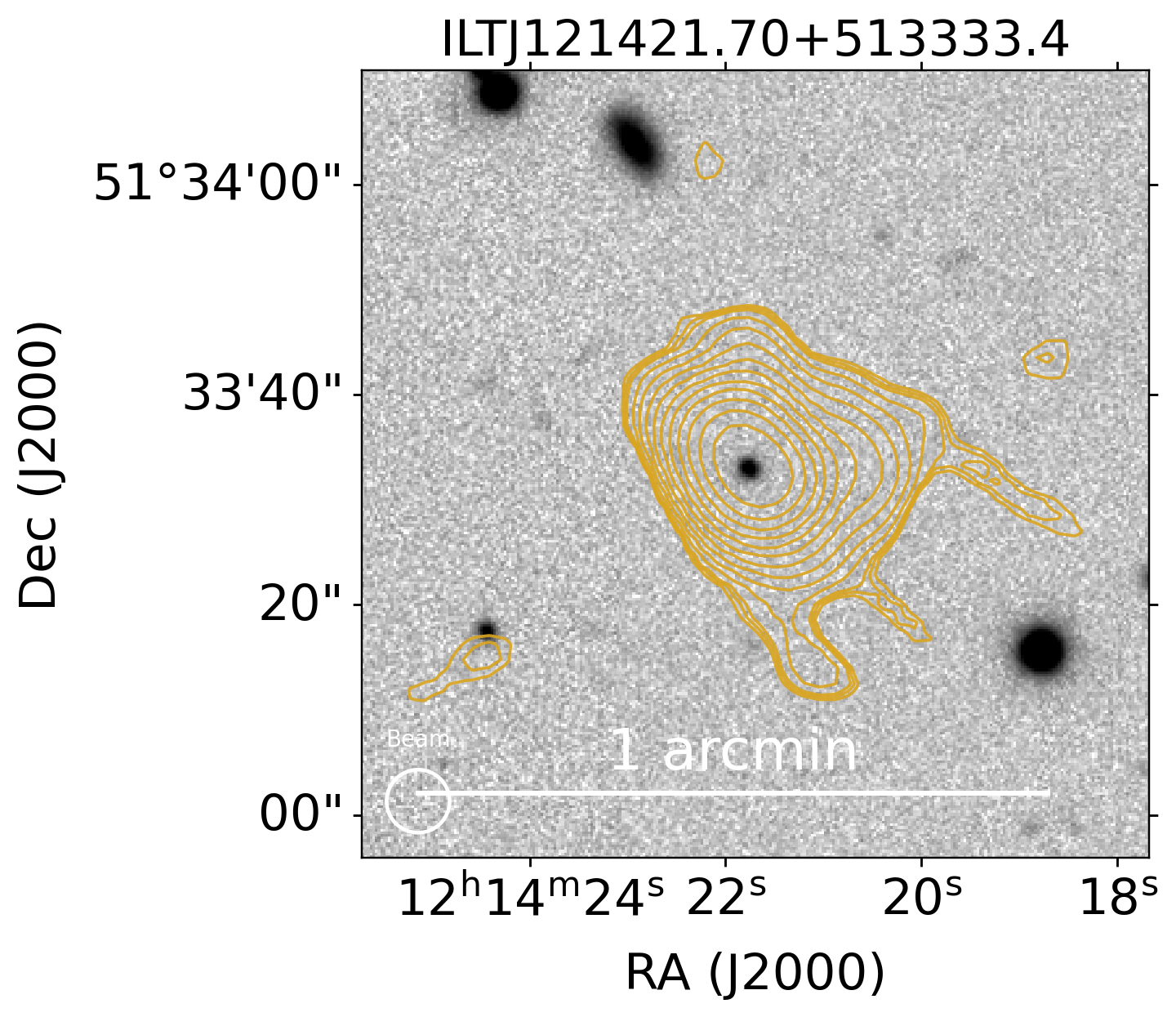}\hfill
  \includegraphics[width=0.188\textwidth]{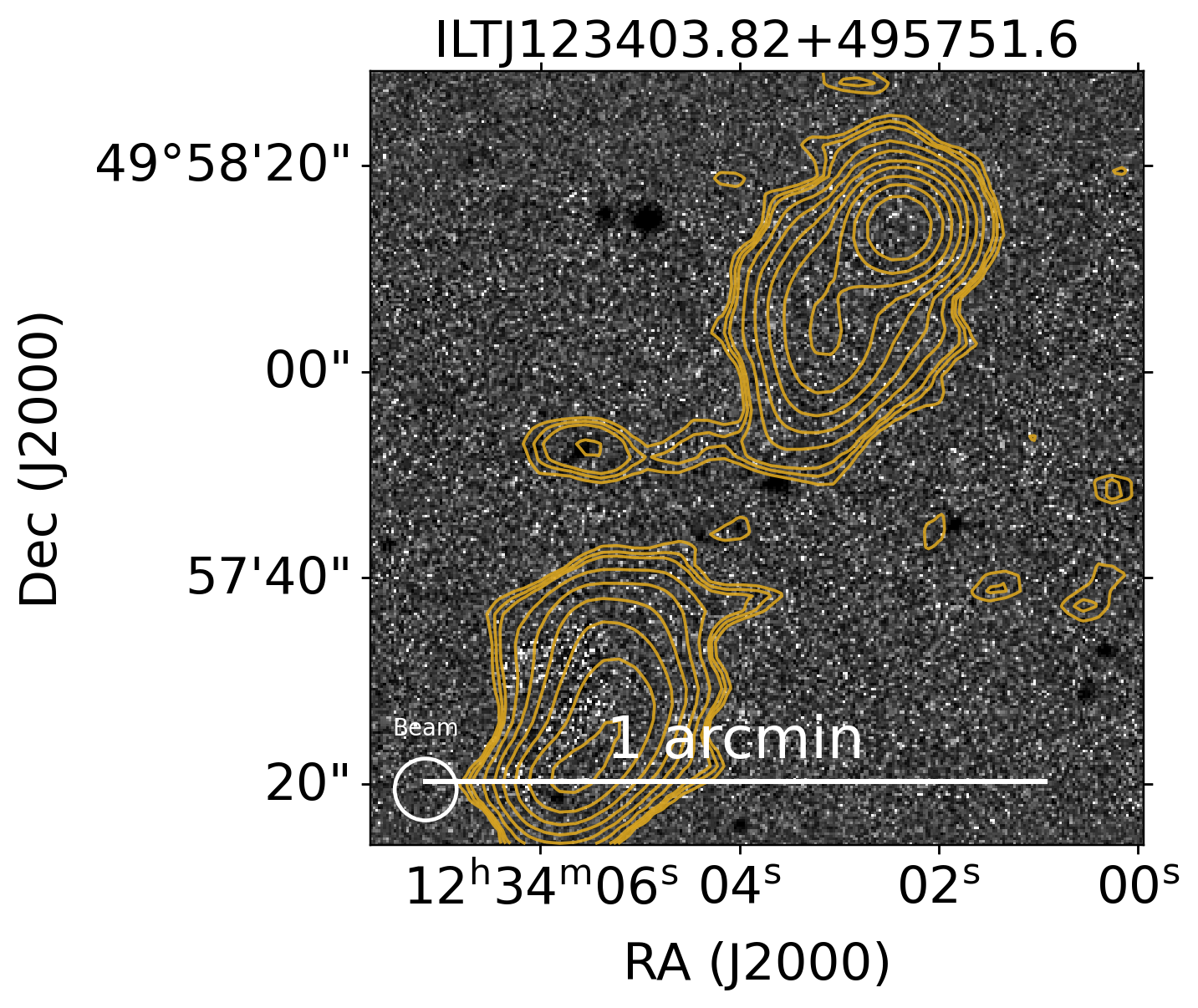}\hfill
  \includegraphics[width=0.188\textwidth]{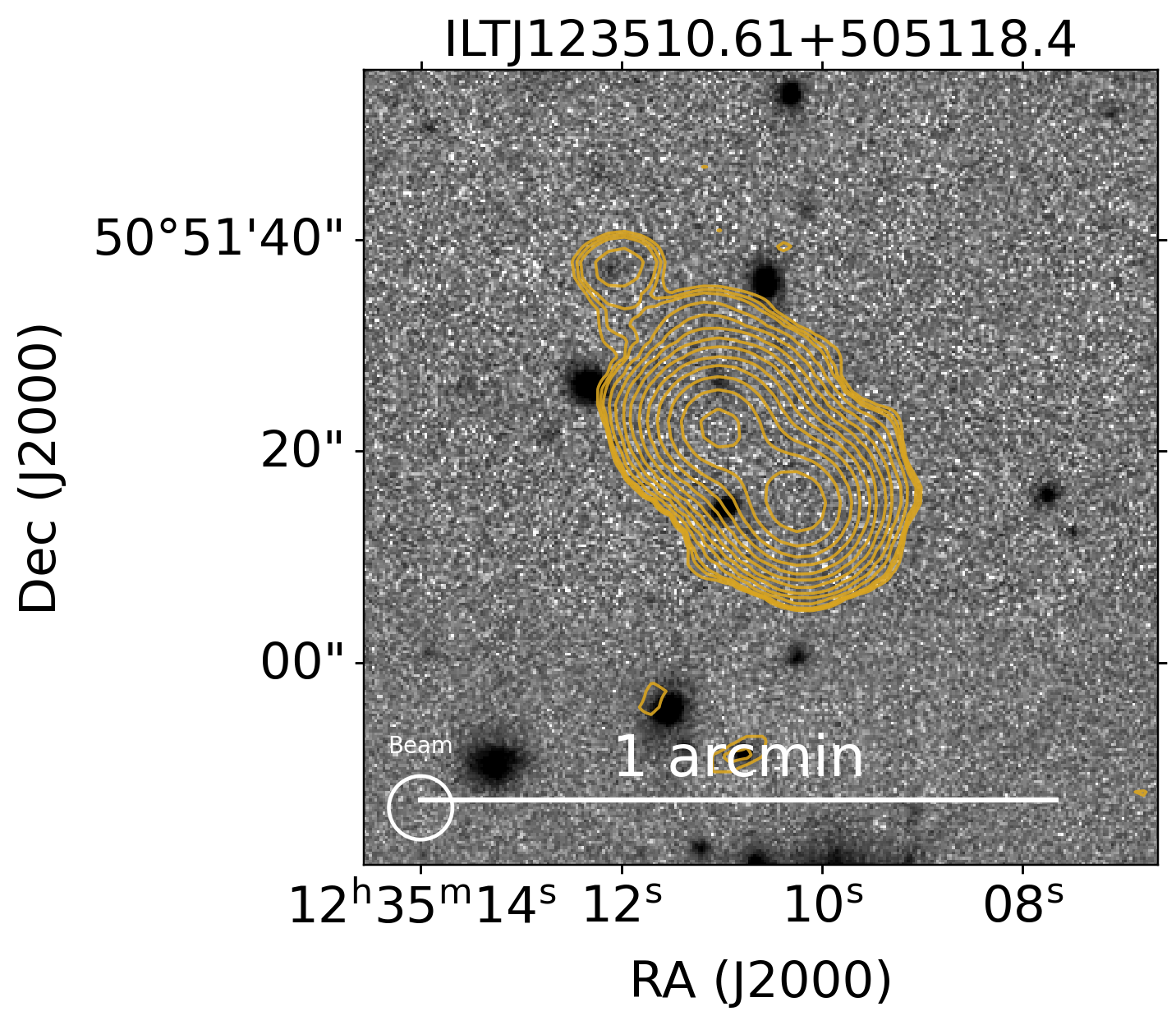}\hfill
  \includegraphics[width=0.188\textwidth]{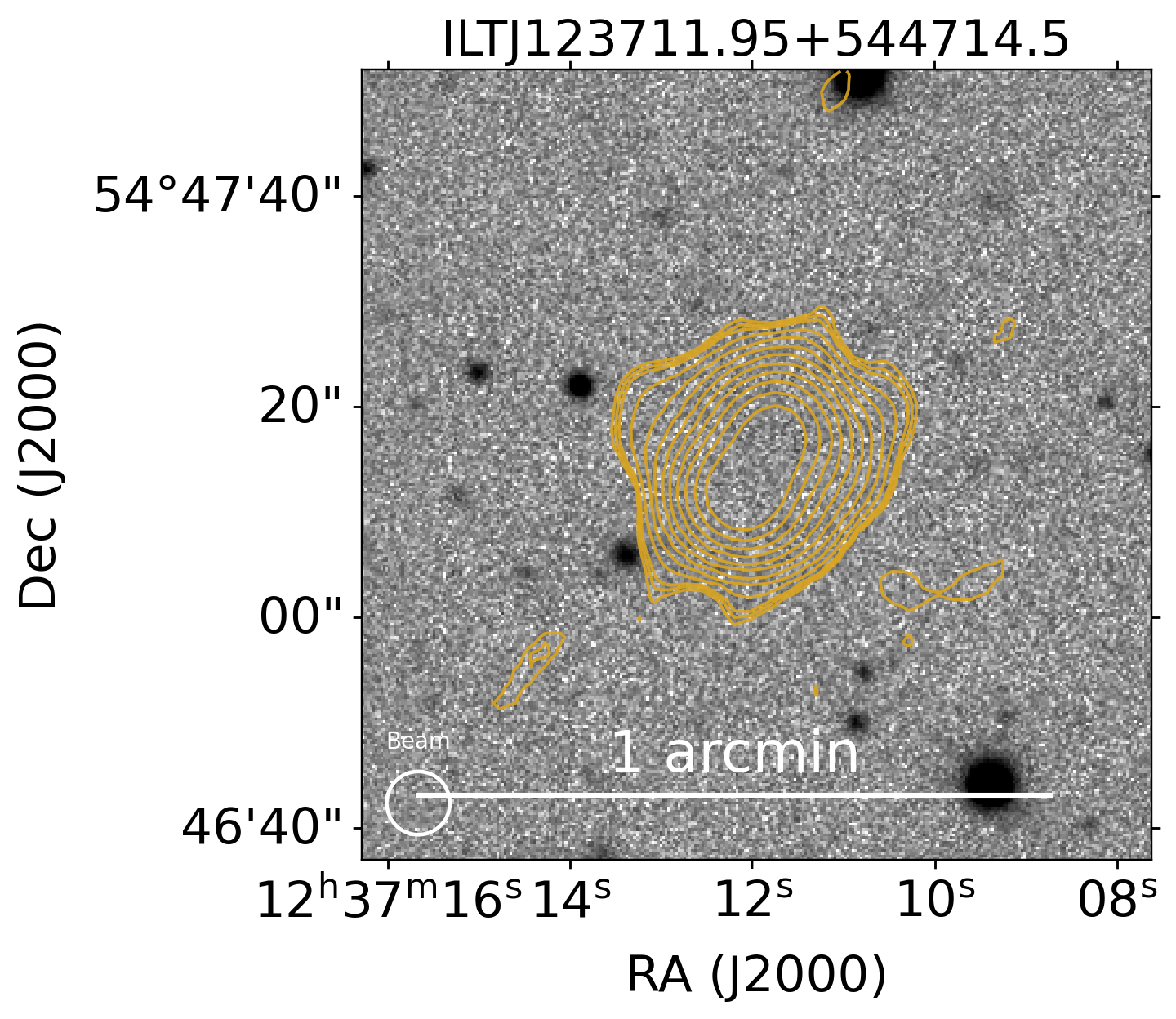}\hfill
  \includegraphics[width=0.188\textwidth]{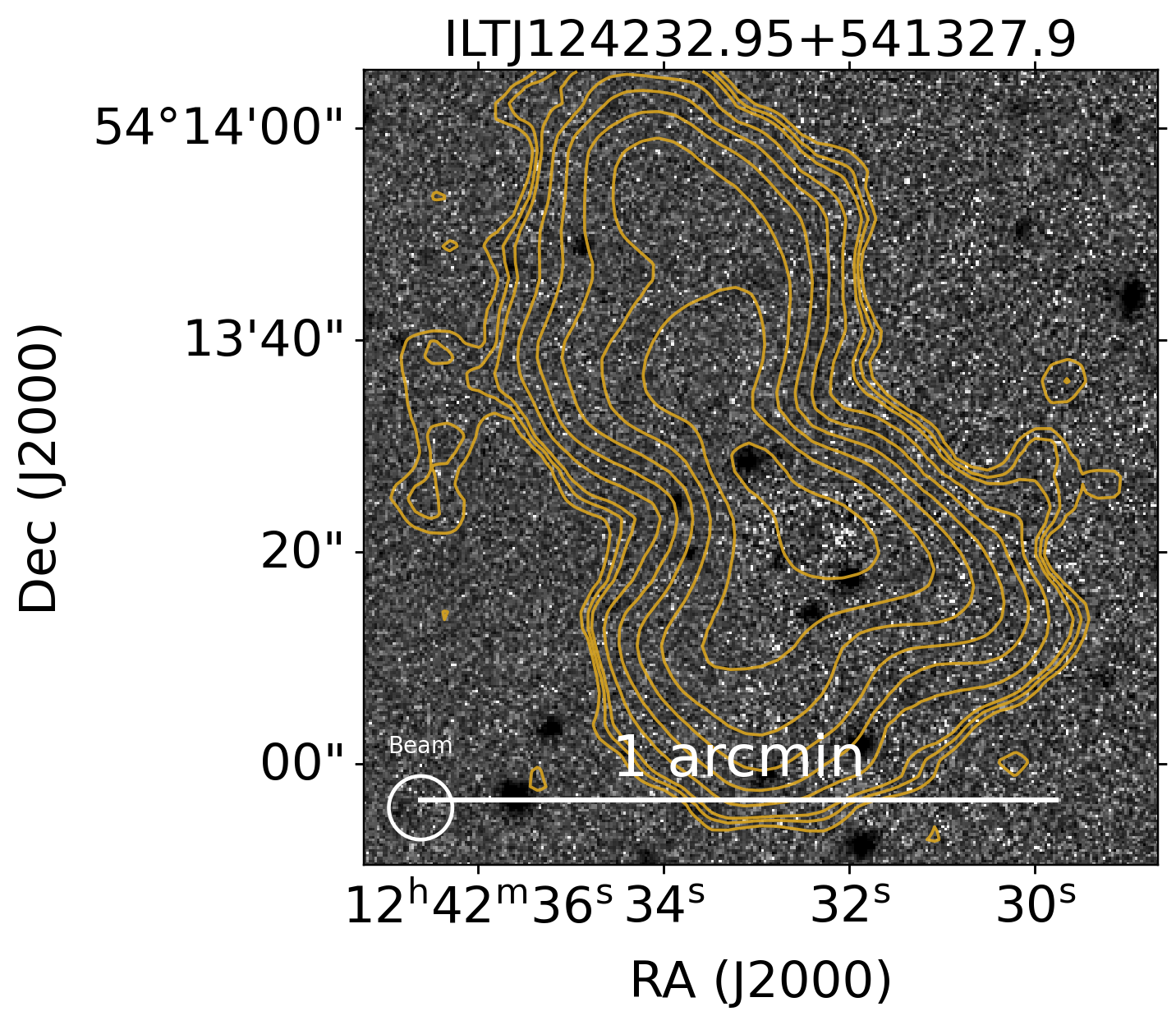}
  
  \par\vspace{1pt}
  
  %---- Row 3 ----%
  \includegraphics[width=0.188\textwidth]{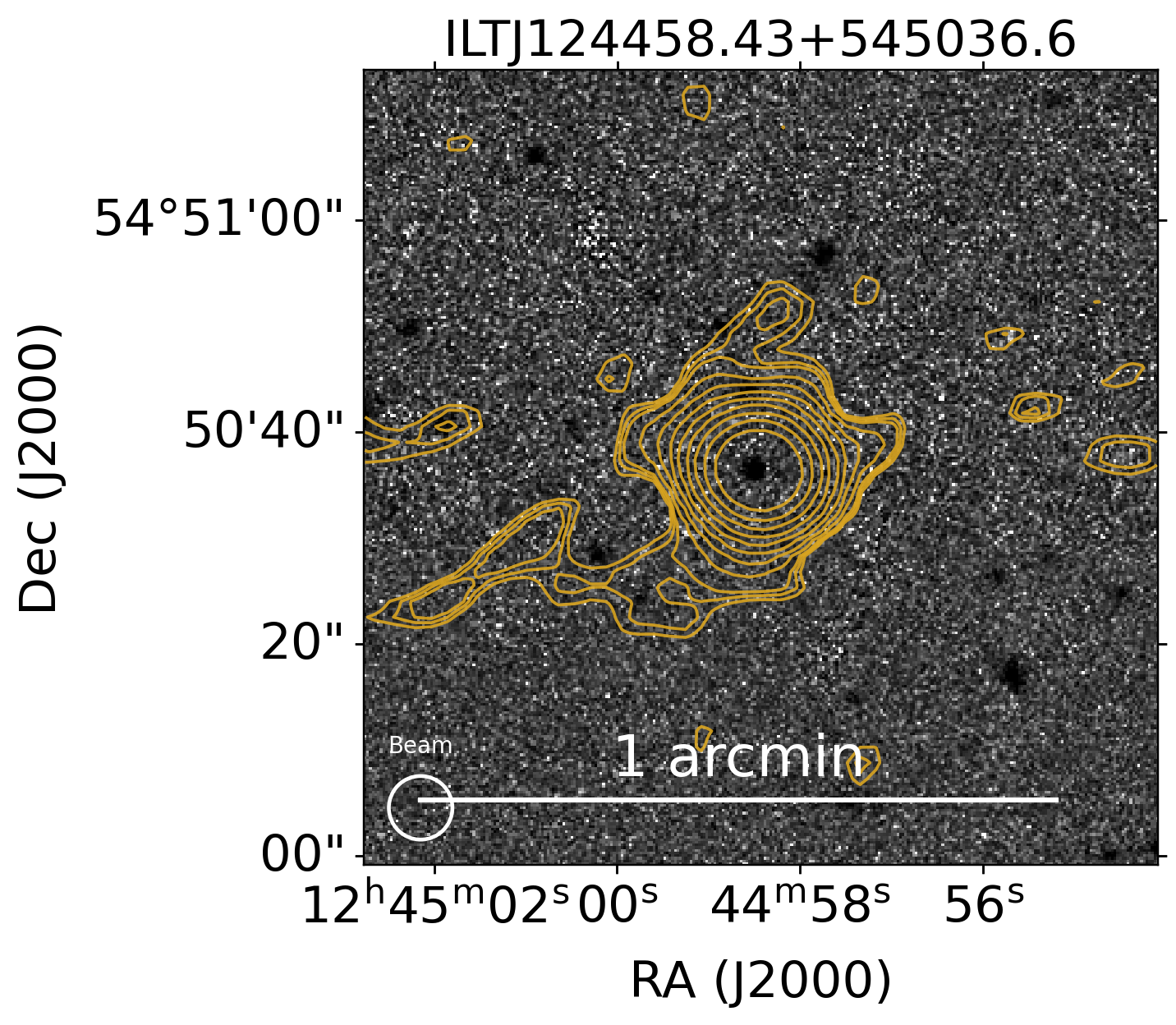}\hfill
  \includegraphics[width=0.188\textwidth]{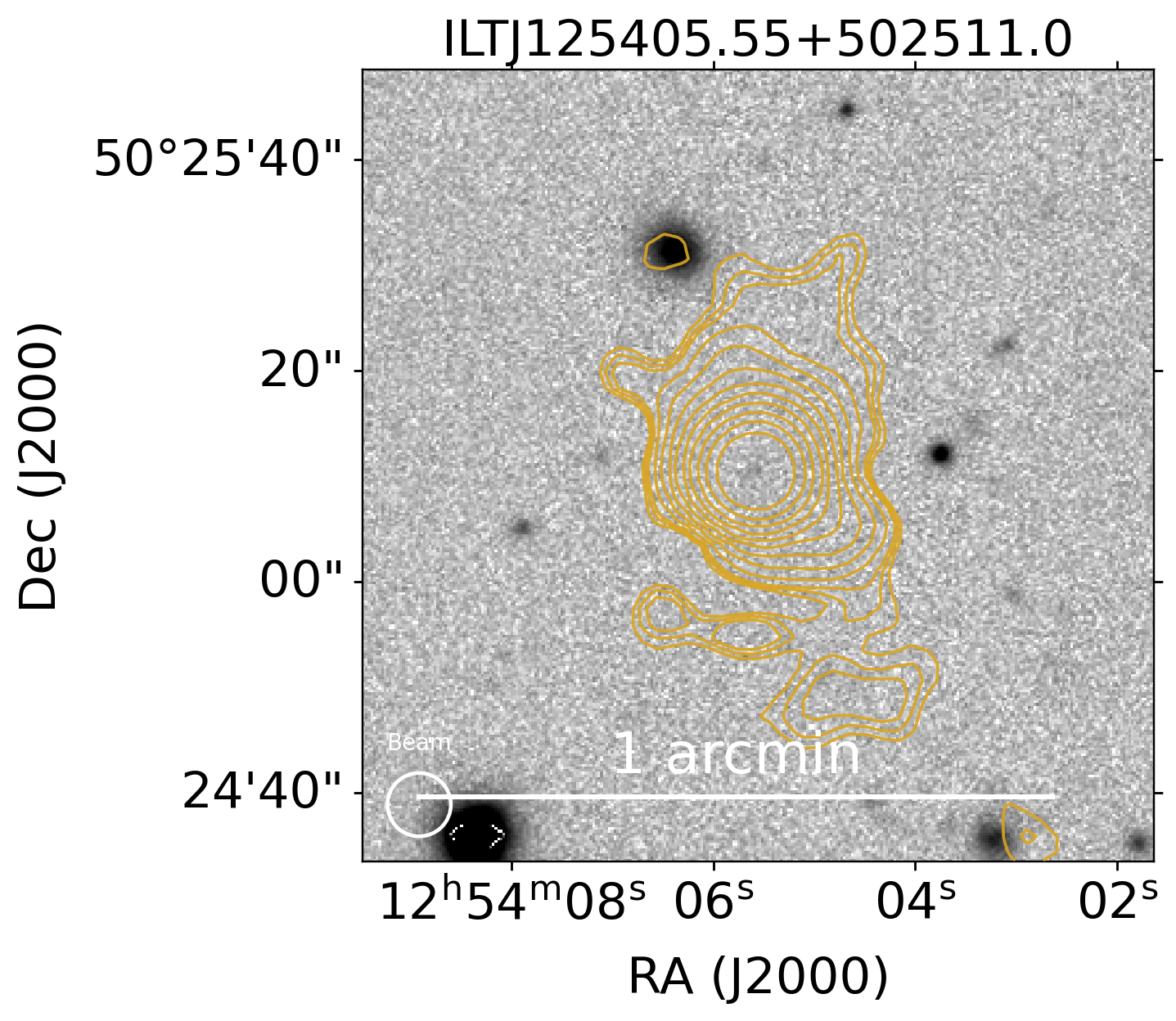}\hfill
  \includegraphics[width=0.188\textwidth]{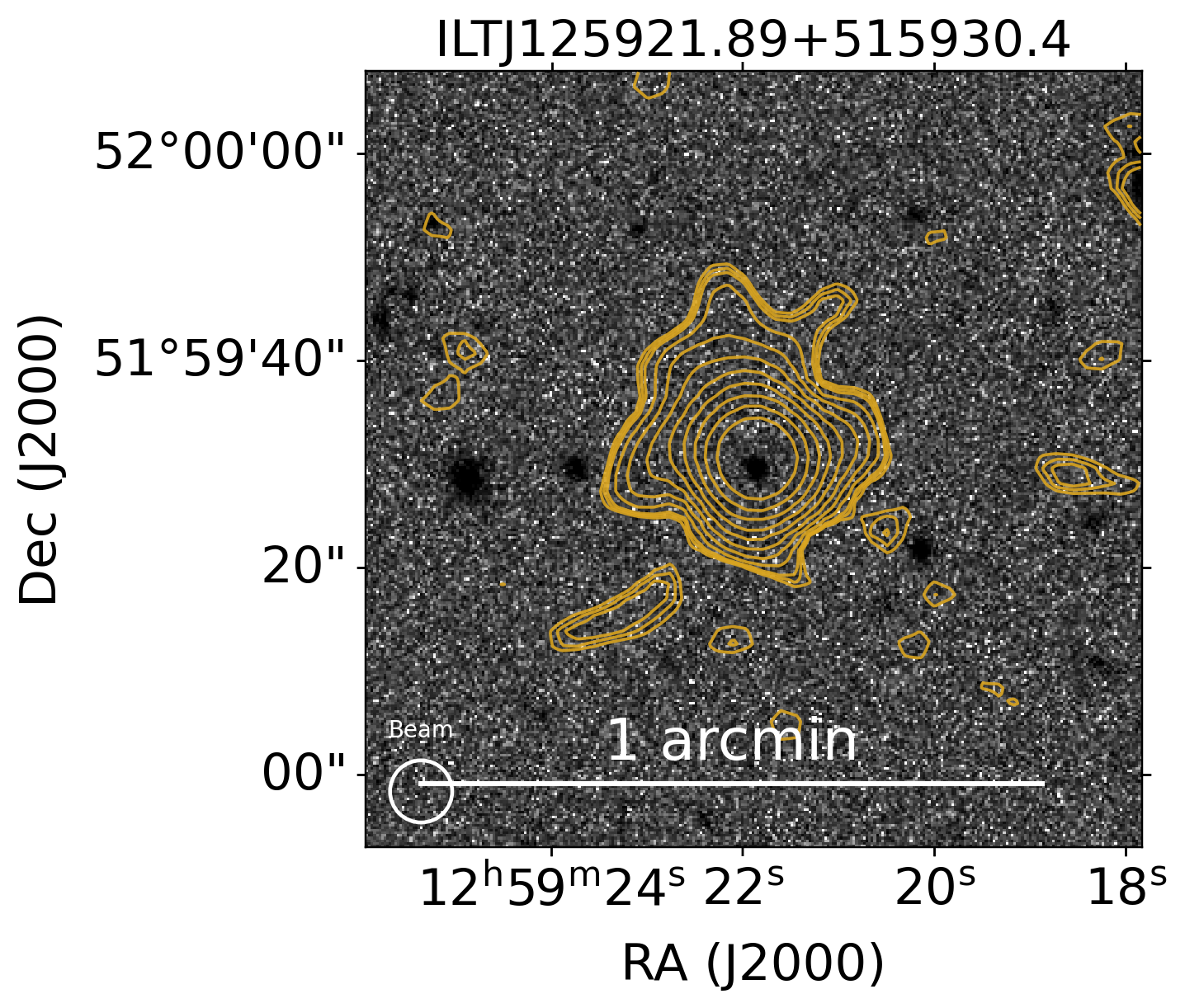}\hfill
  \includegraphics[width=0.188\textwidth]{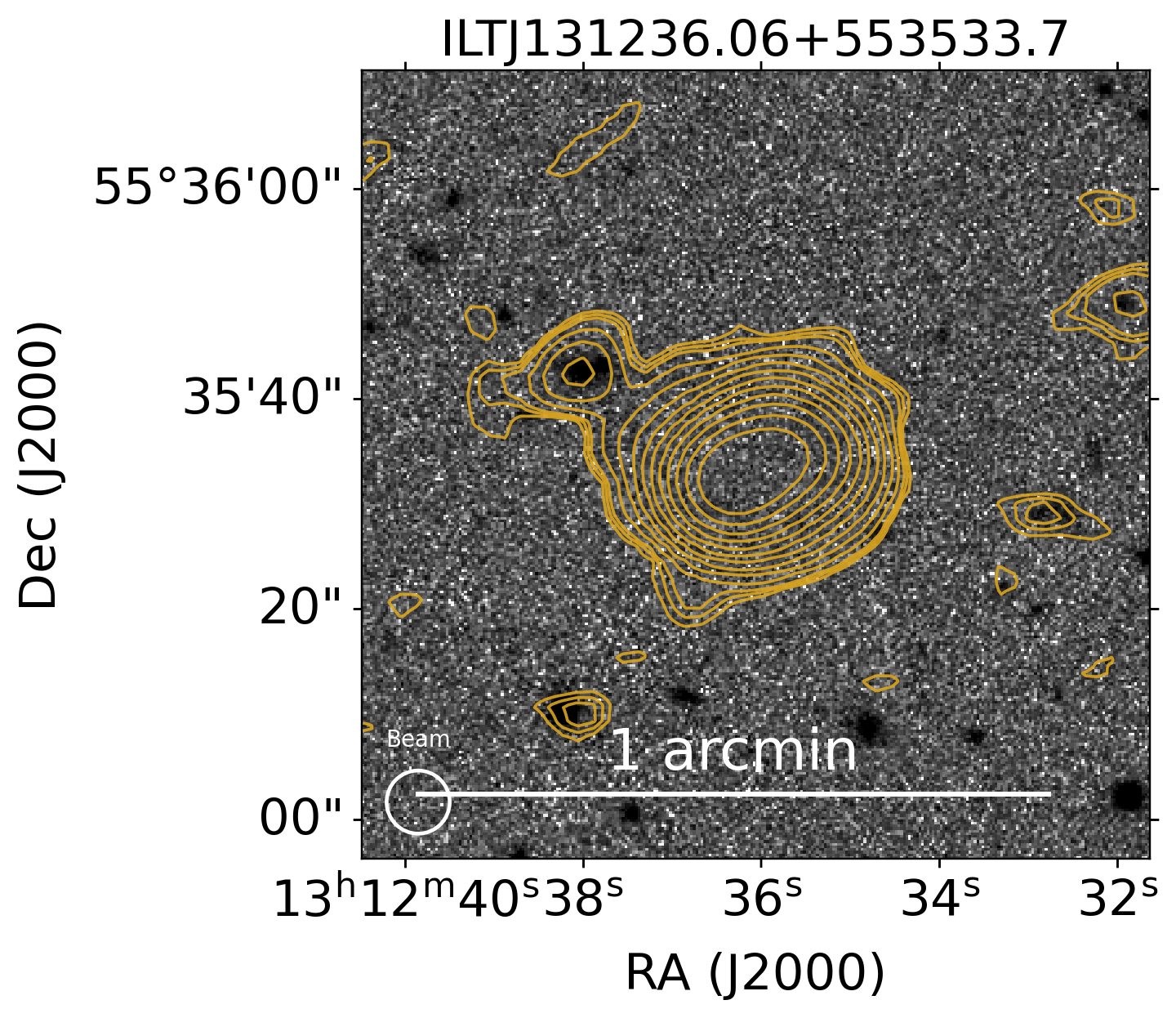}\hfill
  \includegraphics[width=0.188\textwidth]{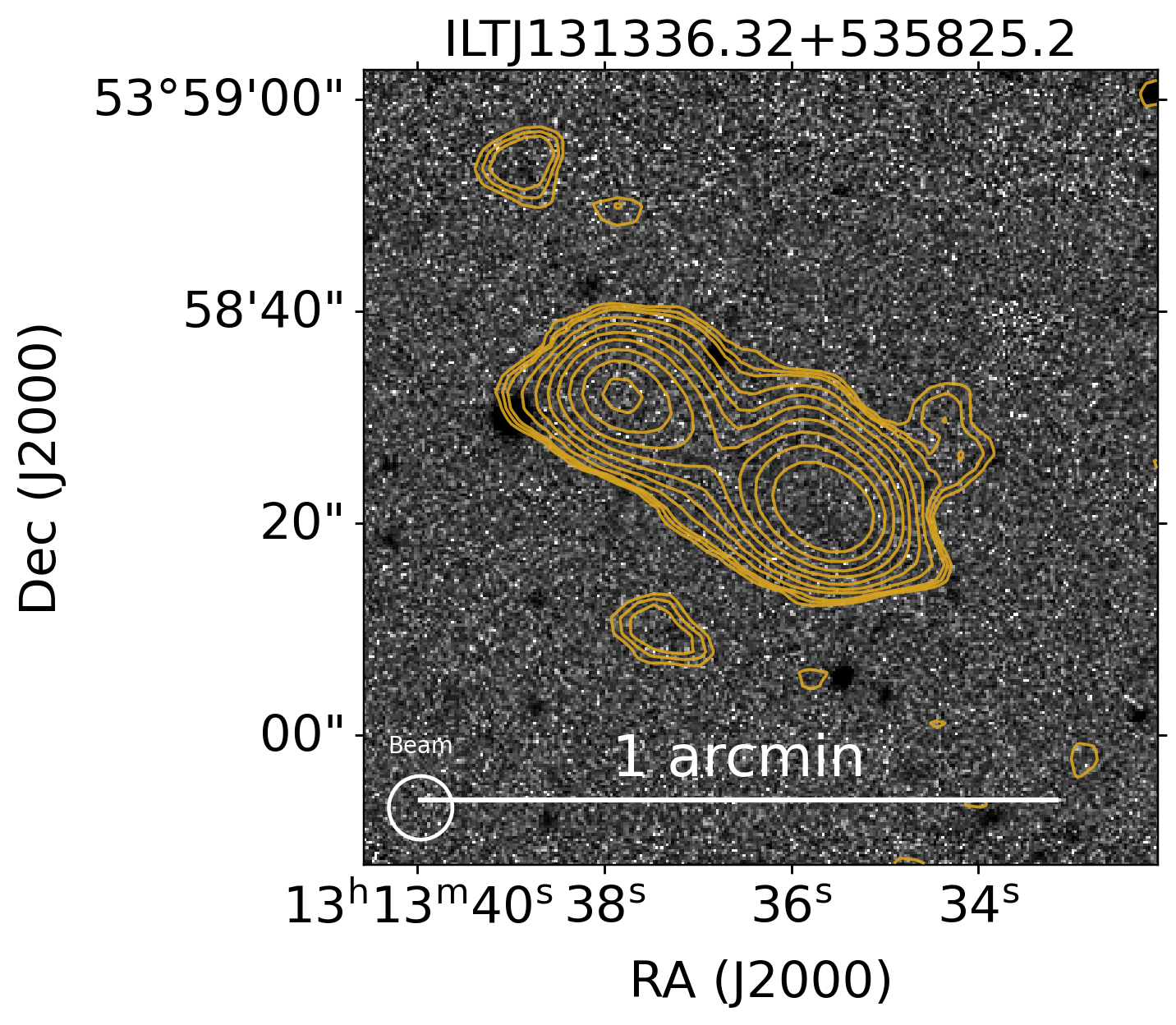}
  
  \par\vspace{1pt}
  
  %---- Row 4 ----%
  \includegraphics[width=0.188\textwidth]{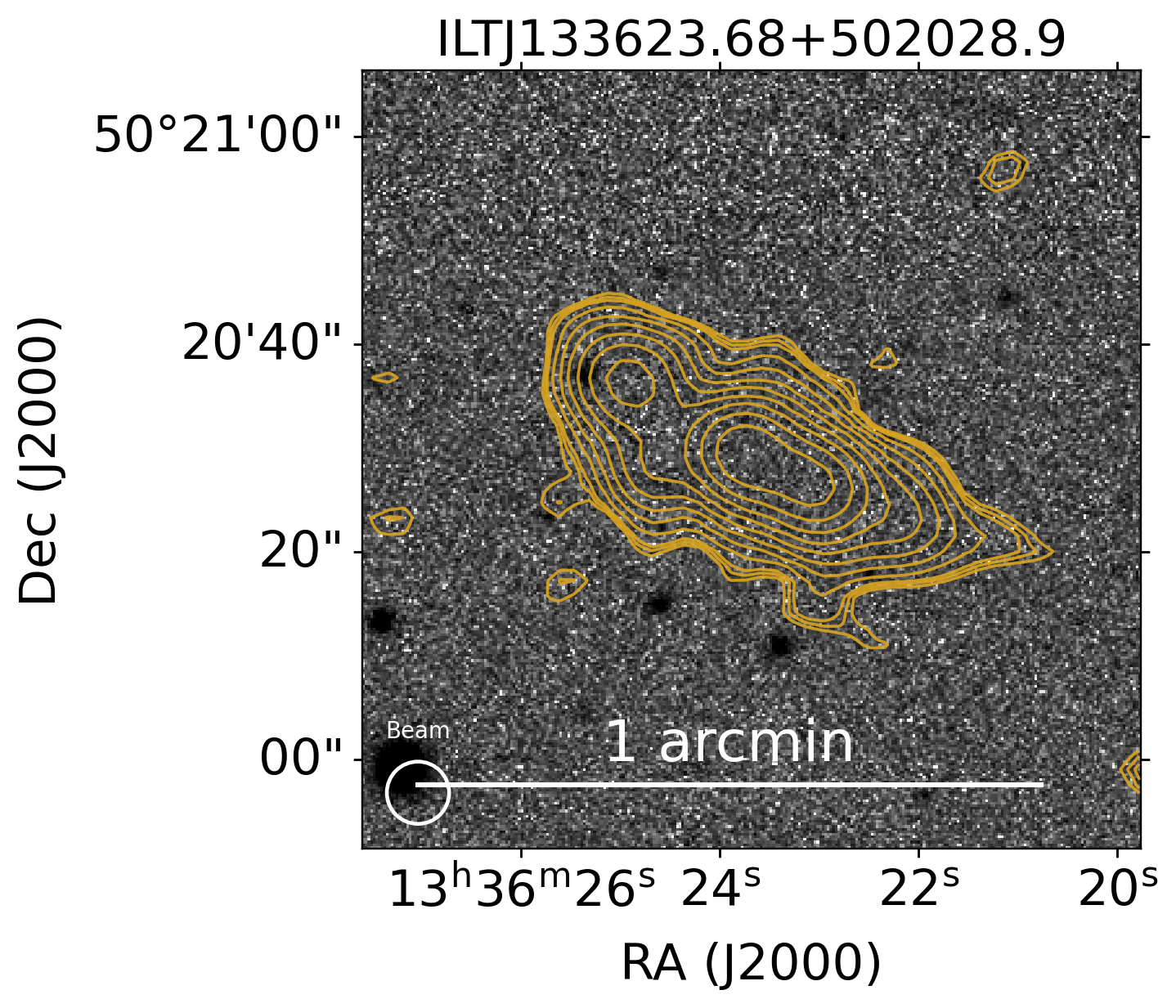}\hfill
  \includegraphics[width=0.188\textwidth]{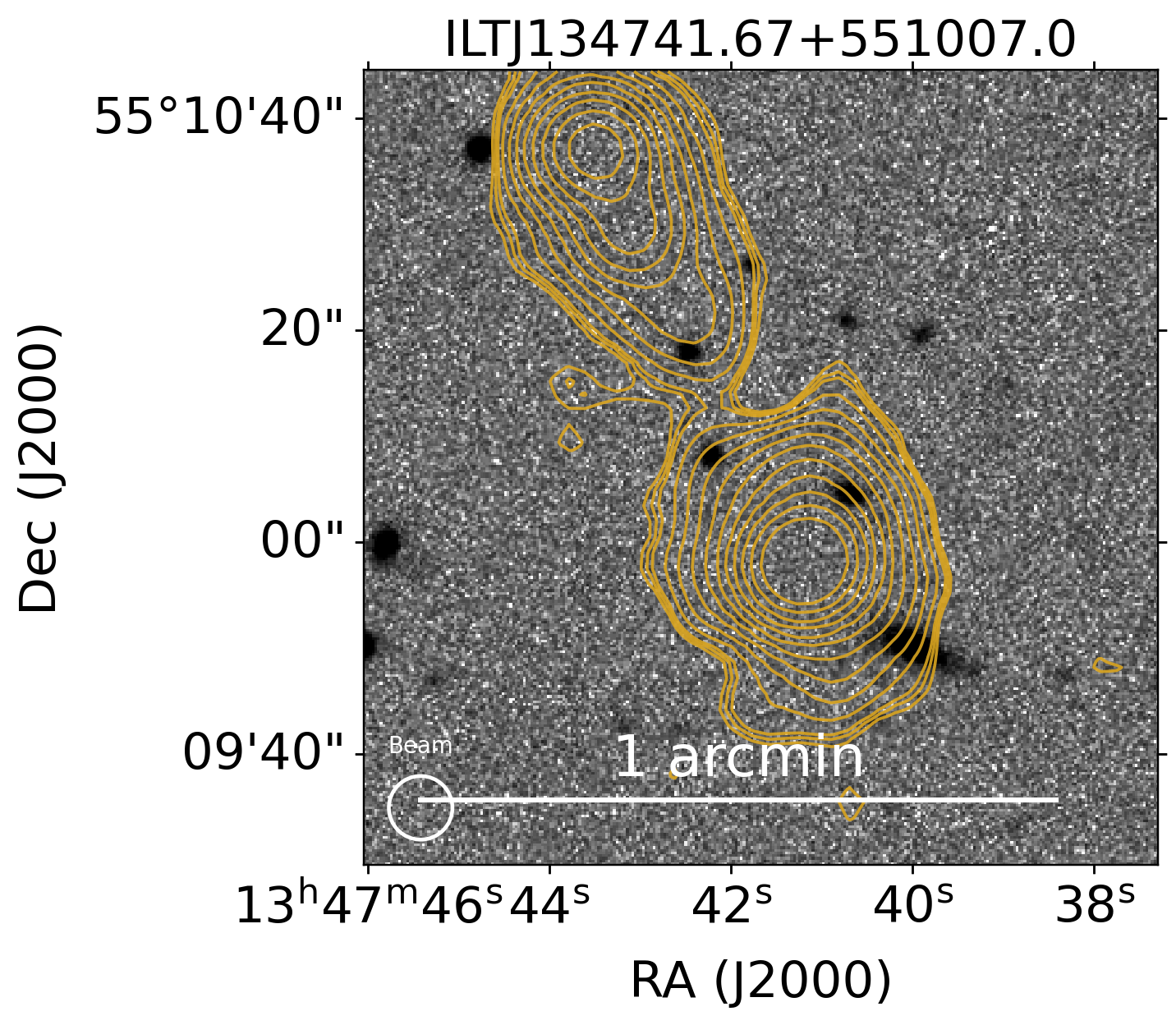}\hfill
  \includegraphics[width=0.188\textwidth]{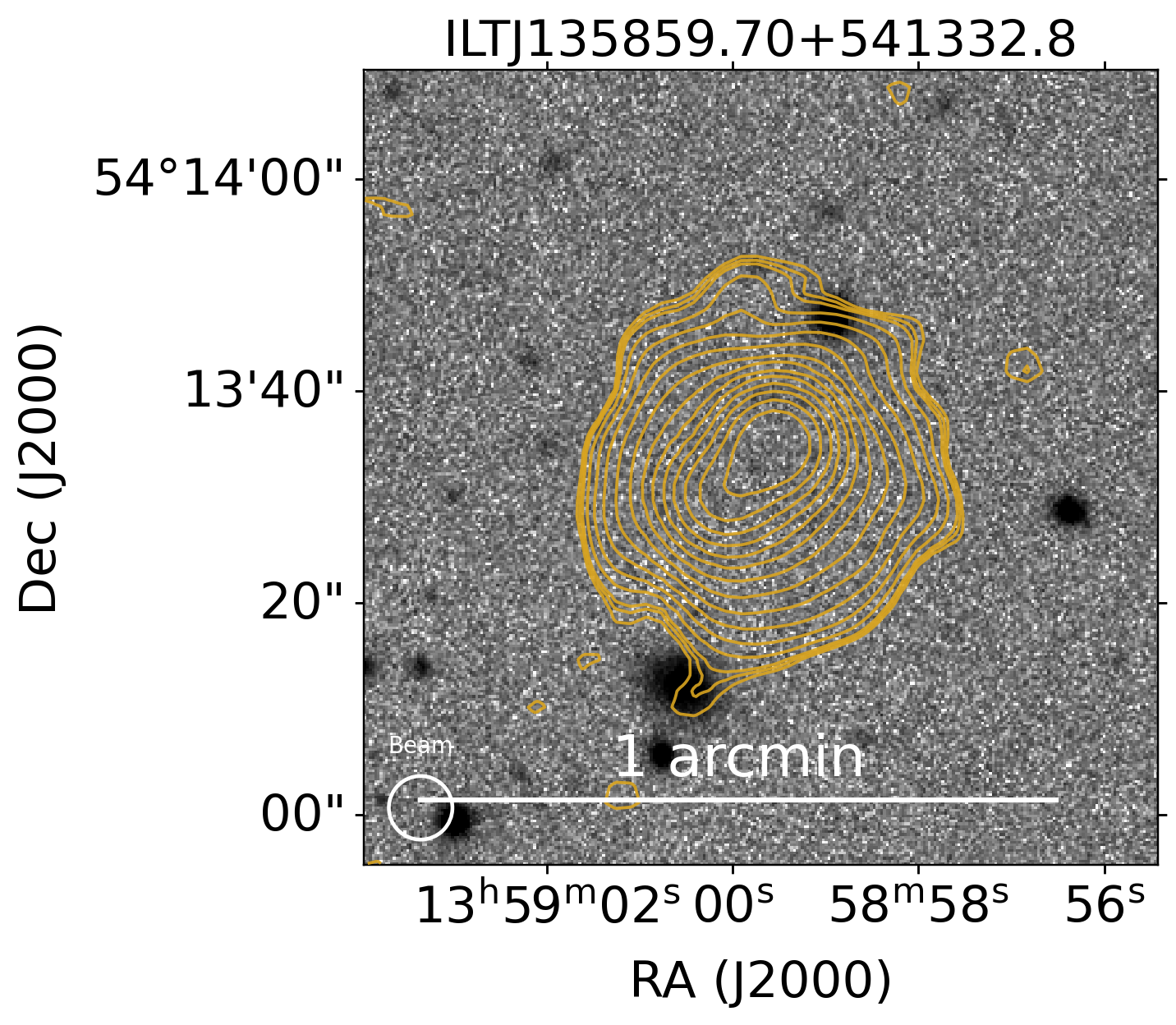}\hfill
  \includegraphics[width=0.188\textwidth]{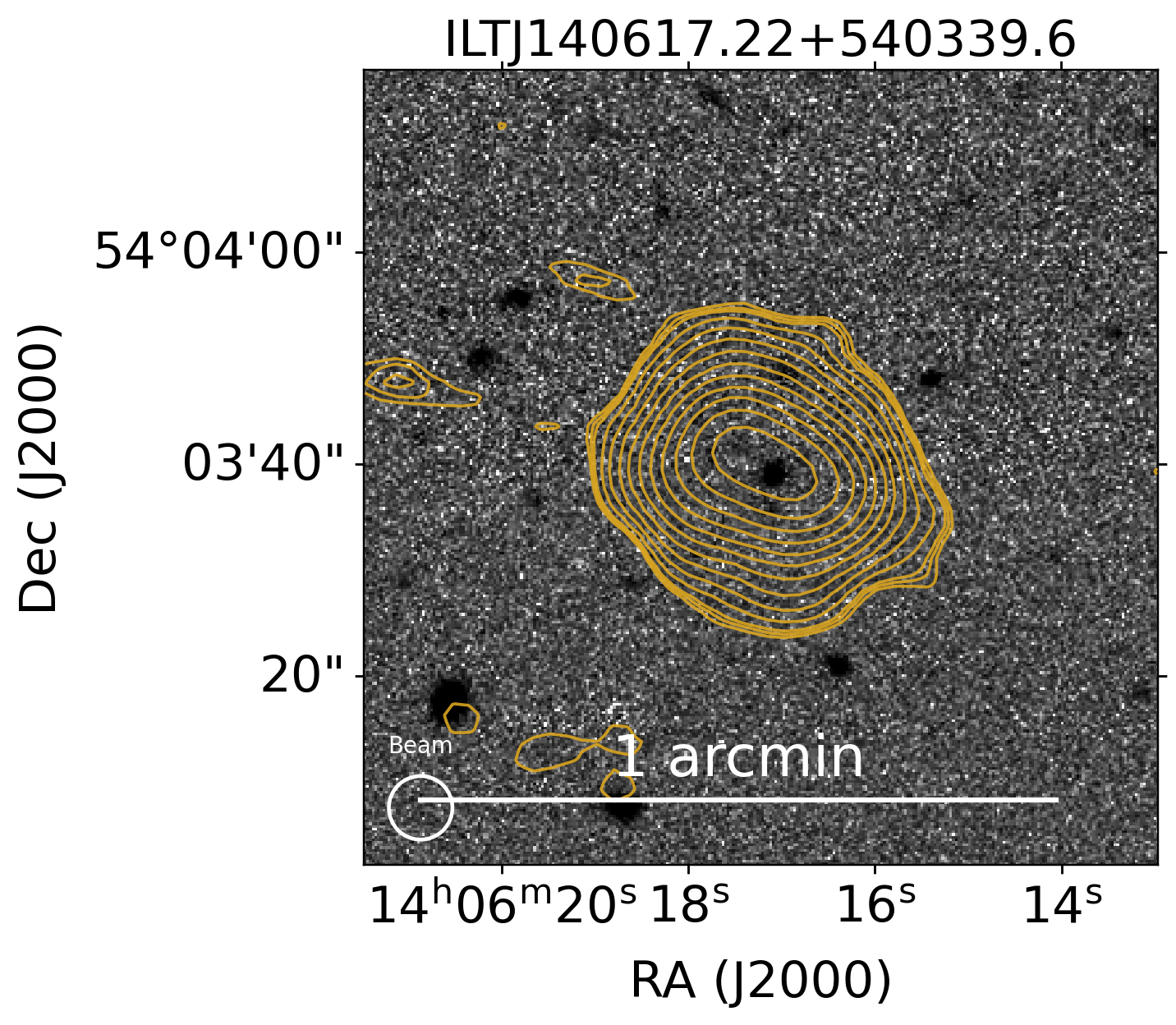}\hfill
  \includegraphics[width=0.188\textwidth]{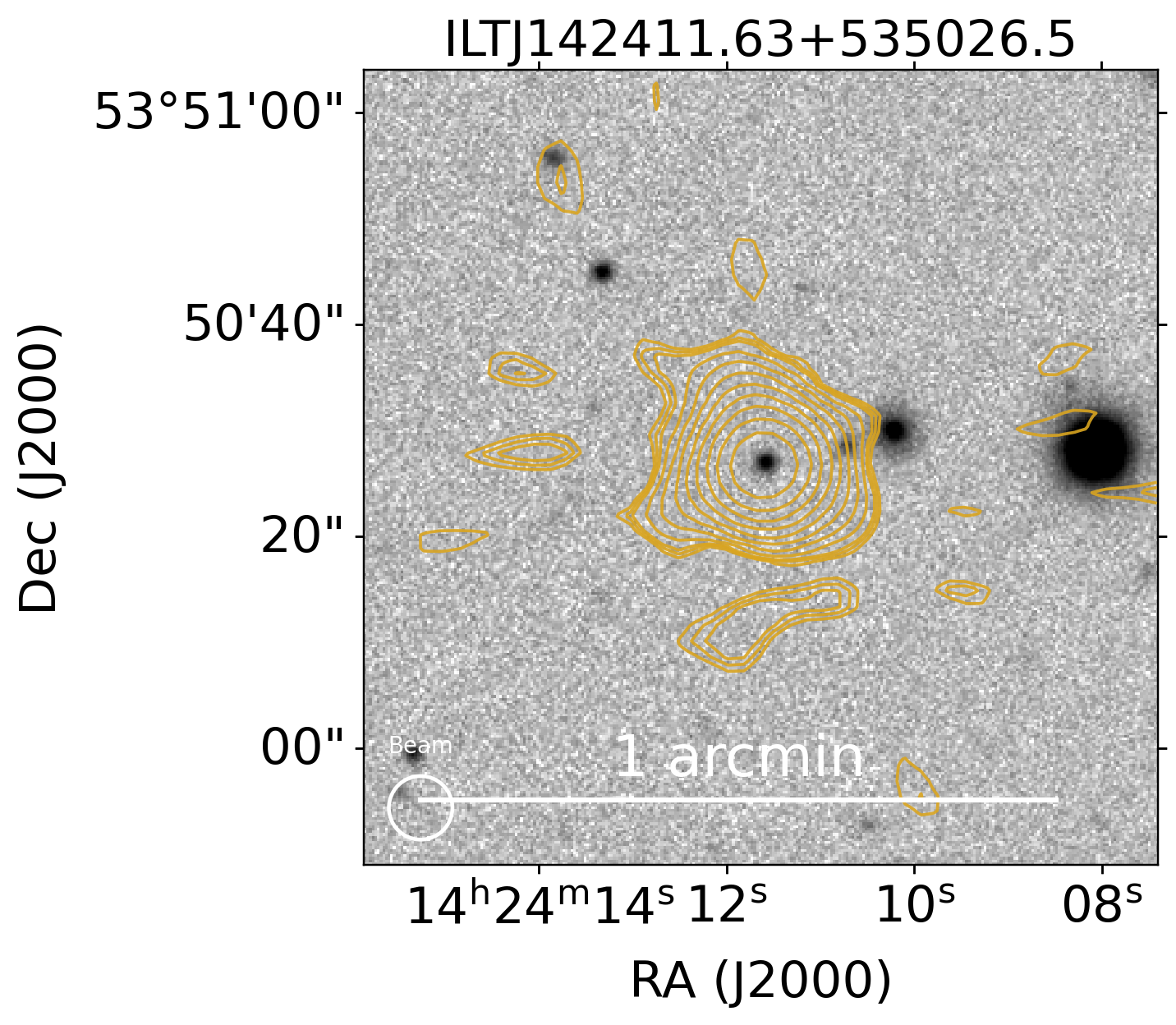}
  
  \par\vspace{1pt}
  
  %---- Row 5 ----%
  \includegraphics[width=0.188\textwidth]{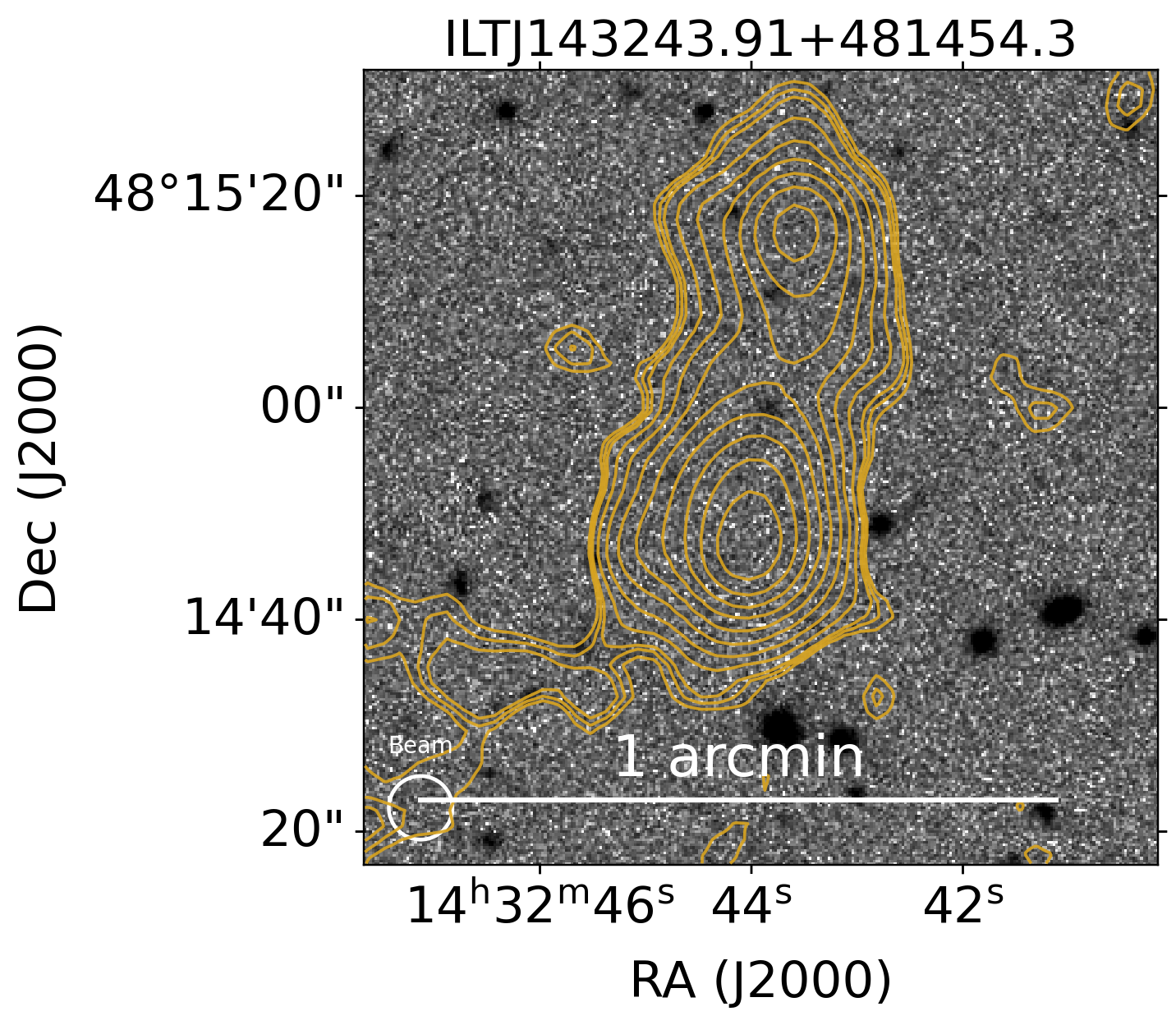}\hfill
  \includegraphics[width=0.188\textwidth]{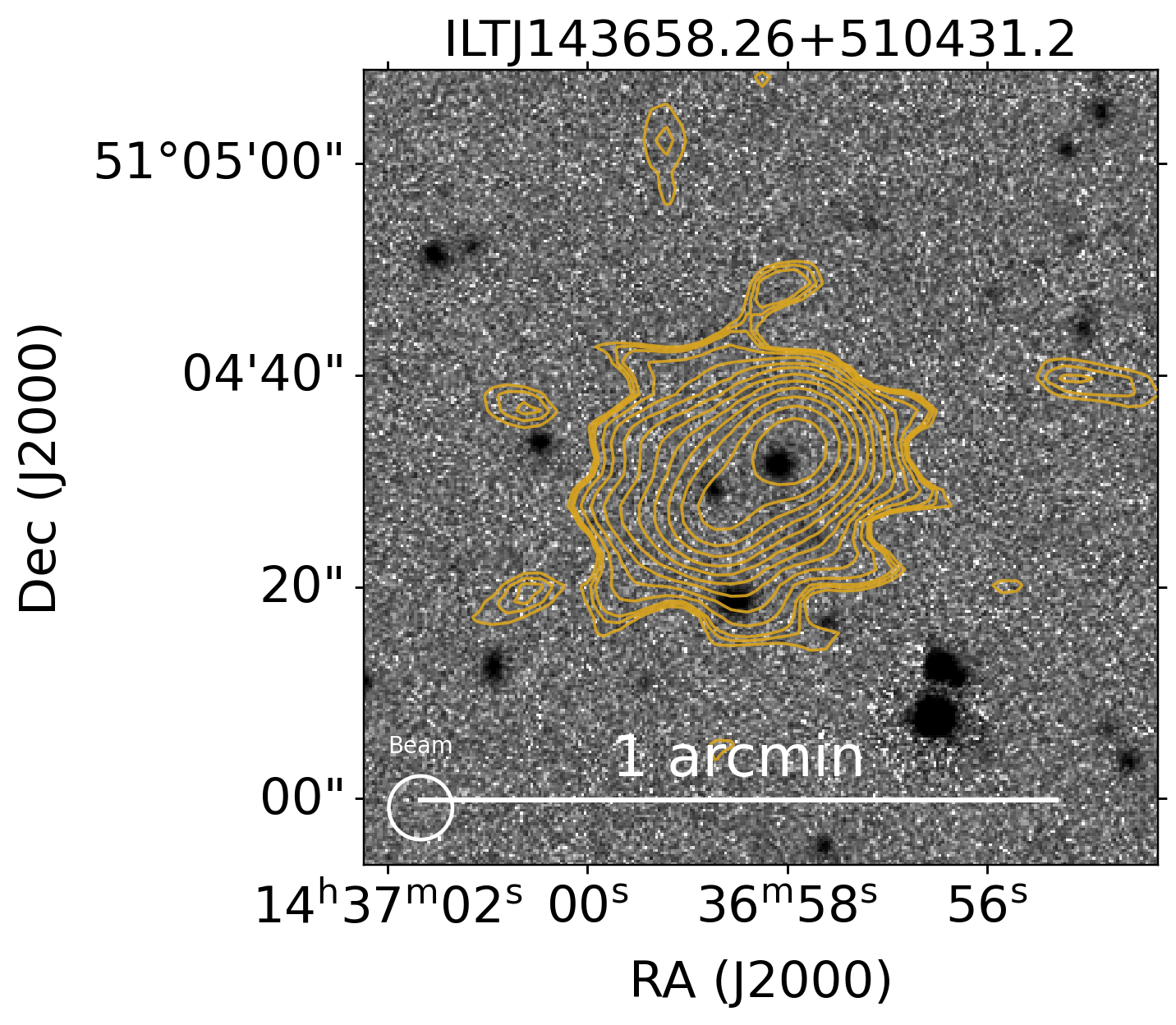}\hfill
  \includegraphics[width=0.188\textwidth]{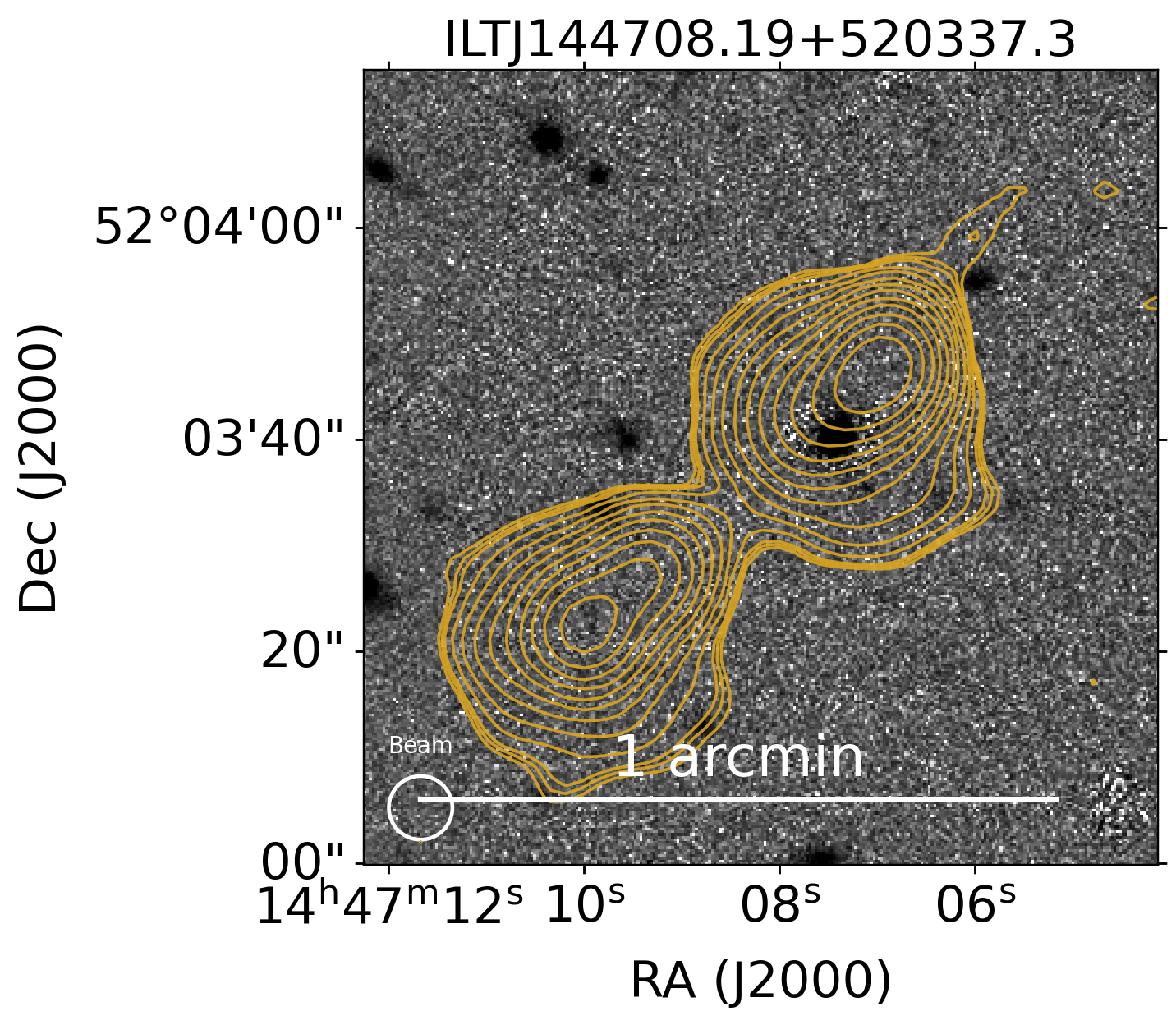}
  
  \caption{Background optical images are from the DESI Legacy Imaging Surveys \citep{Dey2019AJ....157..168D}.
The yellow contours show LOFAR 144 MHz radio emission.
The LOFAR beam size is indicated by the white circle in the lower left corner of each panel,
and the white bar represents a scale of 1 arc-min.}
  \label{fig:extended_sources}
\end{figure*}

In Fig. \ref{fig:extended_sources}, we show the optical images overlaid with the LOFAR intensity maps of 23 resolved radio sources in this work.

\newpage
\section{Additional Table}

In Table \ref{tab:extended}, we summarize the physical properties of 23 resolved radio sources in this work.

\begin{table*}[!ht]
\centering
\caption{Properties of resolved radio sources.}
\label{tab:extended}
\begin{tabular}{lcccccccc}
\hline\hline
Name & RA & Dec & Redshift & FWHM & $L_{\rm Ly\alpha}$ & LAS & Size (Radio) & Size (Ly$\alpha$) \\
 & (deg) & (deg) & & (km\,s$^{-1}$) & ($10^{43}$\,erg\,s$^{-1}$) & (arcsec) & (kpc) & (c kpc) \\
\hline
ILTJ143658.26+510431.2 & 219.2428 & 51.0754 & 2.162 & 2080 & 7.30 & 64.68 & 536 & -- \\
ILTJ105416.58+512326.6 & 163.5691 & 51.3907 & 2.335 & 2636 & 8.26 & 57.19 & 468 & -- \\
ILTJ125921.89+515930.4 & 194.8412 & 51.9918 & 2.067 & 1292 & 8.02 & 48.21 & 402 & -- \\
ILTJ123510.61+505118.4 & 188.7942 & 50.8551 & 2.487 & 735 & 2.18 & 25.82 & 209 & -- \\
ILTJ115937.73+503408.9 & 179.9072 & 50.5692 & 2.568 & 961 & 1.73 & 39.40 & 316 & -- \\
ILTJ121421.70+513333.4 & 183.5904 & 51.5593 & 3.294 & 1842 & 7.08 & 15.41 & 115 & -- \\
ILTJ124458.43+545036.6 & 191.2434 & 54.8435 & 2.202 & 3306 & 6.76 & 46.39 & 383 & -- \\
ILTJ131236.06+553533.7 & 198.1502 & 55.5927 & 3.080 & 578 & 2.65 & 39.43 & 301 & 33 \\
ILTJ142411.63+535026.5 & 216.0485 & 53.8407 & 2.769 & 3242 & 4.16 & 111.39 & 878 & -- \\
ILTJ113805.44+543009.5 & 174.5227 & 54.5027 & 2.084 & 3785 & 3.52 & 23.81 & 198 & -- \\
ILTJ134741.67+551007.0 & 206.9236 & 55.1686 & 2.296 & 1190 & 1.48 & 60.23 & 494 & -- \\
ILTJ131336.32+535825.2 & 198.4014 & 53.9737 & 2.302 & 602 & 1.87 & 46.07 & 378 & 24 \\
ILTJ140617.22+540339.6 & 211.5718 & 54.0610 & 1.959 & 1289 & 7.09 & 20.77 & 174 & -- \\
ILTJ121125.40+535717.6 & 182.8558 & 53.9549 & 2.030 & 1133 & 2.01 & 35.37 & 296 & 26 \\
ILTJ135859.70+541332.8 & 209.7487 & 54.2258 & 2.597 & 1395 & 17.53 & 18.66 & 149 & 56 \\
ILTJ144708.19+520337.3 & 221.7841 & 52.0604 & 2.059 & 1123 & 1.26 & 75.88 & 633 & -- \\
ILTJ124232.95+541327.9 & 190.6373 & 54.2244 & 2.538 & 1223 & 1.03 & 75.85 & 610 & 22 \\
ILTJ123711.95+544714.5 & 189.2998 & 54.7874 & 2.285 & 780 & 1.54 & 17.13 & 141 & -- \\
ILTJ143243.91+481454.3 & 218.1829 & 48.2484 & 2.855 & 395 & 0.68 & 55.27 & 432 & 15 \\
ILTJ123403.82+495751.6 & 188.5159 & 49.9643 & 2.556 & 469 & 1.65 & 80.23 & 644 & 33 \\
ILTJ133623.68+502028.9 & 204.0987 & 50.3414 & 2.535 & 845 & 3.54 & 37.09 & 298 & 40 \\
ILTJ125405.55+502511.0 & 193.5231 & 50.4197 & 1.985 & 677 & 1.54 & 60.00 & 503 & -- \\
ILTJ114733.43+530342.7 & 176.8893 & 53.0619 & 2.025 & 756 & 2.13 & 35.80 & 299 & -- \\
\hline
\end{tabular}
\tablefoot{Properties of radio-resolved `LAE--radio AGN' and `optical AGN with Ly$\alpha$ emission'. 
Columns: (1) source name (LoTSS DR1); (2--3) right ascension and declination (ICRS); 
(4) spectroscopic redshift (HETDEX); (5) Ly$\alpha$ line FWHM; 
(6) Ly$\alpha$ luminosity; 
(7) largest angular size, LAS (arcsec) from the LoTSS DR2 optical counterpart catalogue \citep{Hardcastle2023A&A...678A.151H}; 
(8) radio physical size computed from the LAS using the angular-diameter distance; 
(9) Ly$\alpha$ size (co-moving kpc) from \cite{Erin_inprep}. 
Dashes indicate that a Ly$\alpha$ size is not available.}

    \label{tab:extended}
\end{table*}
\end{appendix}
\end{document}